\documentclass[epsfig,psfig]{aa}

\usepackage{graphicx}
\usepackage{txfonts}
\begin{document}

\title{The formation of the [$\alpha$/Fe] radial gradients in the stars of elliptical galaxies}
\titlerunning{[$\alpha$/Fe] radial gradients in ellipticals}

\author{Antonio Pipino$^{1,3}$, Annibale D'Ercole$^2$, and Francesca Matteucci$^3$}
\authorrunning{A. Pipino et al.}

\institute{
$^1$ Astrophysics, University of Oxford, Denys Wilkinson Building, Keble Road, Oxford OX1 3RH, U.K.\\
$^2$ INAF-Osservatorio Astronomico di Bologna, via Ranzani 1, 40127 Bologna, Italy\\
$^3$ Dipartimento di Astronomia, Universit\`a di Trieste, Via G.B. Tiepolo 11, 34100 Trieste, Italy}

%\author{Cristina Chiappini}
%\affil{INAF-Osservatorio Astronomico di Trieste, Via G.B. Tiepolo 11, 34100 Trieste, Italy}
\date{Accepted,
      Received }

%\maketitle

\abstract{}
{The scope of this paper is two-fold: i) to test and improve our previous models of an outside-in
formation for the majority of ellipticals in the context of the
SN-driven wind scenario, by means of a careful study of gas
inflows/outflows; ii) to explain the observed slopes, either positive
or negative, in the radial gradient of the mean stellar [$\alpha$/Fe],
and their apparent lack of any correlation with all the other
observables.}
{In order to pursue these goals we present a new class of hydrodynamical simulations for the formation
of single elliptical galaxies in which we implement detailed
prescriptions for the chemical evolution of H, He, O and Fe.}
{We find that all the
models which predict chemical properties (such as the
central mass-weighted abundance ratios, the colours as well as the
[$<Fe/H>$] gradient) within the observed ranges for a typical elliptical, also exhibit a
variety of gradients in the [$<\alpha/Fe>$] ratio, in agreement with
the observations (namely positive, null or negative). 
All these models undergo an outside-in formation,
in the sense that star formation stops earlier in the outermost than
in the innermost regions, owing to the onset of a galactic wind.
We find that the predicted variety of the gradients in the [$<\alpha/Fe>$]  
ratio can be explained by physical processes, generally not taken into 
account in simple chemical evolution models, such as  
radial flows coupled with different 
initial conditions  for the galactic proto-cloud.
The typical [$<Z/H>$]
gradients predicted by our models 
have a slope of -0.3 dex per decade variation in radius,
consistent with the mean values of several observational samples. However, we
also find a quite extreme model in which this slope is -0.5 dex per
decade, thus explaining some recent data on gradients in
ellipticals.} 
{We can safely conclude that the history of star formation is fundamental for the creation of abundance gradients in ellipticals but that radial flows with different velocity in conjunction with the duration and efficiency of star formation in different galactic regions are responsible for the gradients in the  [$<\alpha/Fe>$] ratios.}

\maketitle

%\keywords{}

\section{Introduction}

In this paper we exploit the radial variations in the chemical
properties of the Composite Stellar Populations (CSPs) inhabiting
elliptical galaxies in order to gain new insights into the mechanism
of galaxy formation.  Theoretical investigations show, in fact, that
these properties strongly vary as a function of either the duration or
the intensity of the star formation (SF), as well as of the infall
history at each radius.  A useful tool to understand the complex issue of
galaxy formation is represented by the study of the radial gradients
in either the mean abundance ratios or in the mean metallicity in the
stars. There is general consensus on the fact that the observed
increase of line-strength indices such as the $Mg_2$ and the $<Fe>$
(e.g. Carollo et al., 1993; Davies et al., 1993; Trager et al., 2000a)
and the reddening of the colours (e.g. Peletier et al. 1990) toward
the centre of elliptical galaxies, should be interpreted as an
increase of the mean metallicity of the underlying stellar
populations.  In particular, the existence of possible trends of the
gradient slopes with the galactic mass not only could favour a
specific galaxy formation scenario, but also it might tell us about the
degree of uniformity of this process.  
Davies et al. (1993) did not find any correlation linking the
gradients to the mass or the $Mg_2$ of the galaxies, whereas Carollo
et al. (1993) claimed a bimodal trend with mass, in which the $Mg_2$
gradient grows with mass below a certain galactic mass 
($\sim 10^{11} M_{\odot}$) and becomes
flatter in more massive  ellipticals. On the other hand, Gonzalez $\&$
Gorgas (1996) found that the gradient correlates better with the
central value of $Mg_2$ than with any other global
parameter. Kobayashi \& Arimoto (1999) analysed a compilation of data
in the literature, finding that the metallicity gradients do not
correlate with any physical property of the galaxies, including
central indices and velocity dispersion, as well as mass and B
magnitude. Finally, in a very recent paper, Ogando et al. (2005)
claimed that the relation originally found by Carollo et al. (1993)
for low mass ellipticals, might be extended to very massive spheroids
(see also Forbes et al. 2005).
 
From a theoretical point of view, instead, dissipative collapse models
(Larson 1974, Carlberg, 1984) predicted quite steep gradients which
correlate with galactic mass.  Mergers, on the other hand, are
expected to dilute the gradients (Kobayashi, 2004).  In the framework
of chemical evolution models, Martinelli et al. (1998) suggested that
gradients can arise as a consequence of a more prolonged SF, and thus
stronger chemical enrichment, in the inner zones. In the galactic
core, in fact, the potential well is deeper and the supernovae (SN)
driven wind develops later relative to the most external regions (see
also Carollo et al. 1993).  Similar conclusions were found by Pipino
\& Matteucci (2004, PM04), with a more sophisticated model which takes
also into accont the initial infall of gas plus a galactic wind
triggered by SN activity.  PM04 model predicts a logarithmic slope for
indices such as $Mg_2$ which is very close to typical observed gradients, and, on
the average, seems to be independent from the mass of the galaxies.

Gradients in abundance ratios such as the [$\alpha$/Fe] ratio are in principle 
very important, since we could use them as an clock for the
duration of the SF process in that region (see
Matteucci \& Greggio 1986, Matteucci 1994). However, we will show that
the estimate of the \emph{relative} duration of the star
formation process between two different galactic regions with
similar mean [$\alpha$/Fe] ratios in their stars ([$<\alpha/Fe>$], hereafter), 
is also affected by either
the \emph{local} SF efficiency or  by  (differential) metal-enhanced gas
flows.  This is one of the main novelties of our approach with respect
to our previous work.  A prediction made by the PM04 best model was
that the galaxy should form outside-in with an increase in the [$<\alpha/Fe>$] 
ratio  as a function of the radius.  To date, only a handful of observational 
works inferred the gradients in the [$<\alpha/Fe>$] ratios from the indices
such as $Mg_2$ and $<Fe>$ (Melhert et al. 2003, Annibali et al. 2006, 
Sanchez-Blazquez et al. 2007).
These papers show that the slope in the [$<\alpha/Fe>$] gradient can be
either negative or positive, with a mean value close to zero, and that
it does not correlate with galactic properties.  In other words, they
suggest that there is not a preferred mechanism for the formation of
single galaxies, such as either an outside-in or an
inside-out mechanism, at work.  
A drawback of these studies is that their samples are
relatively small and the variations in the indices had been often
evaluated either well inside one effective radius or by neglecting the
galactic core, thus rendering the compilations of the slopes not
homogeneous. % if the actual gradients do not have a linear change with radius.  
On the other hand, a few recently observed single galaxies
(NGC4697, Mendez et al., 2005; NGC821, Proctor et al., 2005, even though in the latter
the authors use an empirical conversion in order to obtain [O/Fe]), seem to
support PM04's predictions, as shown by Pipino, Matteucci \& Chiappini
(2006, PMC06).  PMC06 also stressed the fact the ellipticals are made
of composite stellar populations (CSPs) with properties changing 
with radius; therefore, it cannot be
taken for granted that the abundance pattern used to build theoretical
SSP and to infer abundance ratios from the line indices really reflect
the actual chemical composition of the stars (see also Serra \& Trager
2006).

Finally, a limitation of the chemical evolution models is that 
gas flows cannot be treated with the same detail of a hydrodynamical
model. This may affect not only the infall history or the development
of the galactic wind, but also hampers an estimate of the role of
possible internal flows on the build-up of the gradients.

The aim of this paper is, therefore, manyfold: \\ i) to test the PM04
prediction of an outside-in formation for the majority of ellipticals
in the context of the SN-driven wind scenario by means of a careful
study of gas inflows/outflows; \\ ii) to improve the PM04 formulation
by means of a detailed treatment of gas dynamics; \\ iii) to show how
the observed variety of slopes in the [$<\alpha/Fe>$] gradients in
stars might be related to the different initial conditions and
reconciled within a quasi-monolithic formation scenario. \\ In this
sense we complete and supersede the work of Kobayashi (2004), who,
with SPH models, studied only the metallicity gradients and found that
nearly half of ellipticals have a pure monolithic origin, while the
other half had undergone mergers during their life.  In order to do
that, we couple a simplified chemical evolution scheme with a
hydrodynamical code (Bedogni \& D'Ercole, 1986; Ciotti et al. 1991)
presented in Section 2, whereas our model results will be discussed in
Section 3, 4 and 5; we summarise our main conclusions in Section 6.

\section{The model}

\subsection{Hydrodynamics}

We adopted a one-dimensional hydrodynamical model which follows the
time evolution of the density of mass ($\rho$), momentum ($m$) and
internal energy ($\varepsilon$) of a galaxy, under the assumption of spherical
symmetry.  In order to solve the equation of hydrodynamics with source
term we made use of the code presented in Ciotti et al. (1991), which
is an improved version of the Bedogni \& D'Ercole (1986) Eulerian,
second-order, upwind integration scheme (see their Appendix), to which
we refer the reader for a thorough description of both the set of
equations and their solutions.  Here we report the gas-dynamical
equations:
\begin{equation}
{\partial\rho\over\partial t}+\nabla\cdot(\rho {{u}})=
\alpha\rho_* -\Psi,
\end{equation}
\begin{equation}
{\partial\varrho^i\over\partial t}+\nabla\cdot(\varrho^i {{u}})=
\alpha^i\rho_*-\Psi \varrho^i /\rho	,
\end{equation}
\begin{equation}
{\partial {{m}}\over\partial t}+\nabla\cdot({{m}}
{{u}})=\rho{{g}}-(\gamma-1)
\nabla\varepsilon %+ \alpha {\rho_*} {{u}}_*
-\Psi {{u}} ,
\end{equation}
\begin{equation}
{\partial\varepsilon\over\partial t}+\nabla\cdot(\varepsilon {{u}})=
-(\gamma-1)\varepsilon\nabla\cdot{{u}}-L+\alpha\rho_*
\biggl(\epsilon_0+{1\over 2}u^2\biggr)-\Psi \varepsilon /\rho\, .
\end{equation}
\noindent
The parameter $\gamma=5/3$ is the ratio of the specific heats, ${{g}}$ and
${{u}}$ are the gravitational acceleration and the fluid
velocity, respectively. The source terms on the r.h.s. of equations
(1)--(4) describe the injection of total mass and energy in the gas due
to the mass return and energy input from the stars.  
%${{u}}_*$ is the circular velocity of these stars, and
$\alpha(t)=\alpha_*(t)+\alpha_{\rm SNII}(t)+\alpha_{\rm SNIa}(t)$ is
the sum of the specific mass return rates from low-mass stars and SNe of
both Type II and Ia, respectively.  $\epsilon_0$ is the injection
energy per unit mass due to SN explosions (see Sec. 2.2).  $\Psi$ is
the astration term due to SF.  Finally, $L=n_{\rm e}n_{\rm
p}\Lambda(T,Z)$ is the cooling rate per unit volume, where for the
cooling law, $\Lambda(T,Z)$, we adopt the Sutherland \& Dopita (1993)
curves.  This treatment allows us to implement a self-consistent
dependence of the cooling curve on the metallicity (Z) in the present
code.  We do not allow the gas temperature to drop below $10^4$
K. This assumption does not affect the conclusions.

$\varrho^i$ represents the mass density of the $i-th$ element, and
$\alpha^i$ the specific mass return rate for the same element, with
$\sum^N_{i=1} \alpha^i=\alpha$. Basically, eq. (2) represents a
subsystem of four equations which follow the hydrodynamical evolution
of four different ejected elements (namely H, He, O and Fe).  We
divide the grid in 550 zones 10 pc wide in the innermost regions, and
then slightly increasing with a size ratio between adjiacent zones
equal to 1.03.  This choice allows us to properly sample the galaxies
without wasting computational resources on the fraction of the
simulated box at distances comparable to the galactic tidal radius
{ (see Sec. 2.3 for its value). 
At the same time, however, the size of the simulated box is roughly
a factor of 10 larger than the stellar tidal radius.
This is necessary to avoid that possible perturbations at the boundary
affect the galaxy and because we want to have a surrounding medium which
acts as a gas reservoir for the models in which we start from
an initial flat gas density distribution (see Sec. 2 for the model definition.}
We adopted a reflecting boundary condition in the center of the grid
and allowed for an outflow condition in the outermost point.

At every point of the mesh we allow the SF to occur with the following rate:
\begin{equation}
\Psi = \nu  \rho = {\epsilon_{SF} \over max(t_{cool},t_{ff})} \rho \,
\label{sfr}
\end{equation}
where $t_{cool}$ and $t_{ff}$ are the \emph{local} 
cooling and free-fall timescales, respectively,
whereas $\epsilon_{SF}$ is a suitable \emph{SF parameter} 
{which contains all the uncertainties on the timescales of the SF process that
cannot be taken into account in the present modelling}
{ and its value is given \emph{a priori}.
{ In particular, we stress that the adopted parametrization of the SF process might
appear simplistic, although it is a rather standard assumption in many galaxy formation simulation 
where the sub-grid physics cannot be properly modelled.
A more detailed representation should at least discriminate between a cold molecular gas phase
which is actually feeding the SF process, and the hot surrounding medium where the ejecta
from SN are deposited. On the other hand, eq.~\ref{sfr} does not imply that the SF is occurring
in the hot gas phase; in fact, we assume that a suitable fraction proportional to the average
density in the gridpoint forms stars once it had cooled down.
\footnote{Note also that $\Psi \rightarrow 0$ if $t_{cool} \rightarrow \infty$, namely
if the gas is cooling on a very long timescale.}}

$\nu$ gives the speed of the SF process, whereas the \emph{final efficiency},
namely the fraction of gas which has eventually turned into stars, is an output of the model.}
%provided that the gas is not flowing out of the cell.

We assume that the stars do not move from the gridpoint in which they
have been formed.  We are aware that this can be a limitation of the
model, but we prefer this solution than moving the stars in order to
match some pre-defined luminosity profile (as done in, e.g., Friaca \&
Terlevich 1998), because this might artificially affect the resulting
metallicity gradients. Moreover, we expect that the stars will
spend most of their time close to their apocentre.
In order to
ensure that we match the observed mass-to-light ratio for the given potential well,
we stop the SF in a given grid-point only if
the mass density of low-mass stars created at that radius exceeds a
given threshold profile.  The adopted profile is a King distribution,
{ with core radius of 370 pc and a central stellar mass density of $6\times 10^{-21}\rm g\, cm^{-3}$.
Integrating over the whole galactic volume, the above mentioned
limiting profile yields a total stellar mass of $\sim 3\times 10^{11}M_{\odot}$.}
In the next Section we will show that this assumption does not flaw our
simulated galaxies, because the occurrence of a galactic wind, which
halts the SF process, coincides with or occurs even earlier than the
time at which such a threshold profile is attained.

At the beginning the gas is subject only to the Dark Matter (DM) halo 
gravity and to its
own self-gravity; once the SF begins, the gravitational potential due to the
stellar component is self-consistently evaluated.

The DM potential has been evaluated by assuming a distribution inversely
proportional to the square of the radius at large distances (see Silich \& Tenorio-Tagle 1998).
We classify each model according to the size of the DM halo (see next Section).
The adopted core radii for the DM distribution, instead, are reported
in Table 1.

\subsection{Chemical Evolution}

We follow the chemical evolution of only four elements, namely H, He,
O and Fe. This set of elements is good enough to characterize our
simulated elliptical  galaxyfrom the chemical evolution point of view.  In
fact, as shown by the time-delay model (Matteucci \& Greggio, 1986,
see also PMC06), the [$\alpha$/Fe] ratio is a powerful estimator of
the duration of the SF.  Moreover, both the predicted [Fe/H]-mass and
[Z/H]-mass relationships in the stars can be tested against the
observed Colour-Magnitude Relations (hereafter CMRs;e.g. Bower et
al. 1992) and Mass-Metallicity relation (hereafter MMR; e.g. Carollo
et al. 1993).  In order to clarify this point, we recall that the O is
the major contributor to the total metallicity, therefore its
abundance is a good tracer of the metal abundance Z. However,
we stress that we always refer to Z as the sum of the O and Fe mass abundances. 
On the other
hand, the Fe abundance is probably the most commonly used probe of the
metal content in stars, therefore it enables a quick comparison between our
model predictions and the existing literature.
We are aware that in the past literature the majority of the works
used Mg as a proxy for the $\alpha$ elements, as it can be easily
observed in absorption in the optical bands giving rise to the well
known $Mg_2$ and $Mg_b$ Lick indices. It is worth noticing, however,
that the state-of-the-art SSPs libraries (Thomas et al. 2003, Hyun-Chul Lee
\& Worthey, 2006), are computed as functions of the 
\emph{total} $\alpha$-enhancement
and of the total metallicity. Moreover latest observational results (Mehlert et
al. 2003, Annibali et al. 2006 and Sanchez-Blazquez et al. 2007), 
have been translated into theoretical ones by means of these SSPs;
therefore the above authors provide us with radial gradients in [$\alpha$/Fe], instead of [Mg/Fe].
This is why in this paper we focus on the theoretical evolution
of the $\alpha$ elements, and the O is by far the most important.
In any case, we will also present our predictions in the form of indices
and show that we obtain reasonable values
in agreement with observations.
In fact, we will compare our results to recent observational data 
which have been transformed into abundance ratios by means of SSPs computed 
by assuming a global $\alpha$-enhancement.
Finally, on the basis of nucleosynthesis calculations, we expect
O and Mg to evolve in lockstep. This means that
the [O/Fe]=[Mg/Fe]+const equation should hold (in the gas) 
during galactic evolution
(see e.g. Fig. 1 of PM04);
%, especially because we adopt mass-averaged yields (?)for massive stars (see below); 
therefore the predicted slope of
the [$\alpha$/Fe] gradient in the stars should not change if we adopt
either O or Mg  as a proxy for the $\alpha$s.
There might be only an offset in the zero point of, at most, 0.1-0.2 dex
which is within both the obseved scatter and the uncertainties of
the \emph{calibration} used to transform Lick indices into abundance ratios.

The nucleosynthetic products enter the mass conservation equations via several source terms,
according to their stellar origin.
A Salpeter (1955) initial mass function (IMF) constant in time in the
range $0.1-50 M_{\odot}$ is assumed, since PM04 and PMC06 showed that
the majority of the photochemical properties of an elliptical galaxy
can be reproduced with this choice for the IMF.  We adopted the yields
from Iwamoto et al. (1999, and references therein) for both SNIa and
SNII.  The SNIa rate for a SSP formed at a given radius is
calculated assuming the single degenerate scenario and the Matteucci
\& Recchi (2001) Delay Time Distribution (DTD).  The convolution of
this DTD with $\Psi$ over the
galactic volume, gives the total SNIa rate, according to the following
equation (see Greggio 2005):

\begin{equation}
r_{Ia}(t)=k_{\alpha} \int^{min(t, \tau_x)}_{\tau_i}{A (t-\tau) \Psi(t-\tau) 
DTD(\tau) d \tau}
\end{equation}
where $A(t- \tau)$ is the fraction of binary systems which give rise to Type Ia SNe. 
Here we will assume it constant (see Matteucci et al. 2006 for a more detailed
discussion). 
The time $\tau$ is the delay time defined in the range $(\tau_i, \tau_x)$ so that:

\begin{equation}
\int^{\tau_x}_{\tau_i}{DTD( \tau) d \tau}=1
\end{equation}
where $\tau_i$ is the minimum delay time for the occurrence of Type Ia SNe, 
in other words the time at which the first SNe Ia start occurring. We assume, 
for
this new formulation of the SNIa rate that $\tau_i$ is the lifetime of a 
8$M_{\odot}$, while for $\tau_x$, which is the maximum delay time, 
we assume the lifetime of a  $0.8M_{\odot}$.
The DTD gives the likelihood that at a given time a binary system will
explode as a SNIa.
Finally, $k_{\alpha}$ is the number of stars per unit mass in a stellar 
generation and contains the IMF.

According to the adopted model progenitor
and nucleosynthetic yields, each SNIa explosion releases $E_0 =\epsilon_{SN}\, 10^{51}$ erg of
energy and $1.4 M_{\odot}$ of mass (out of which $0.1 M_{\odot}$ of O
and $0.7 M_{\odot}$ of Fe, respectively).  For the sake of simplicity,
we assume that the progenitor of every SNII is a typical
\emph{average} (in the range $10-50 M_{\odot}$) massive star of $18.6
M_{\odot}$, which pollutes the ISM with $\sim 17 M_{\odot}$ of ejecta
during the explosion (out of which $1.8 M_{\odot}$ of O and $0.08
M_{\odot}$ of Fe, respectively).  We recall that single low- and
intermediate-mass stars do not contribute to the production of either
Fe or O. We neglect the fact that they may lock some heavy elements
present in the gas out of which they formed, and restore them
on very long timescales;
therefore single low- and intermediate-mass stars are only responsible for the ejection
of H and He.
Such a simplified scheme has been also tested with our
chemical evolution code (PM04, their model IIb); it leads to relative
changes smaller than the 10\% in the predicted abundance ratios with
respect to the ones predicted with the full solution of the chemical
evolution equations.
%In all the cases in which we want a higher degree of accuraracy for the evaluation of the gas metallicity Z, we use the relation
%$[Z/H]=[Fe/H]+0.94[\alpha/Fe]$ (Tantalo \& Chiosi, 2004).

These quantities, as well as the evolution of single low and intermediate 
mass stars, had been evaluated by adopting 
the stellar lifetimes given by Padovani \& Matteucci (1993).
The solar abundances are taken from Asplund et al. (2005).

We recall that in order to study the mean properties
of the stellar component in ellipticals, we need average quantities
related to the mean abundance pattern of the stars, which, in turn
can allow a comparison with the observed integrated spectra.
To this scope, we recall that, at
a given radius, both real and model galaxies are made of a Composite
Stellar Population (CSP), namely a mixture of several SSPs, differing
in age and chemical composition according to the galactic chemical
enrichment history, weighted with the SF rate.  On the other hand, the
line-strength indices are usually tabulated only for SSPs as functions
of their age, metallicity and (possibly) $\alpha$-enhancement.

{ In particular we make use of the mass-weighted mean stellar metallicity as defined by
Pagel \& Patchett (1975, see also Matteucci 1994): 
\begin{equation}
<Z>={1\over S_f} \int_0^{S_f} Z(S)\,  dS\, ,
\label{PP75original}
\end{equation}
where $S_f$ is the total mass of stars ever born contributing to the 
light at the present time and Z is the metal abundance (by mass) in the gas
out of which an amout of stars $S$ formed.
In practice, we make use of the stellar mass distribution
as a function of Z in order to derive the 
mean metallicity in stars.

One can further adapt eq.~\ref{PP75original} in order
to calculate the mean O/Fe ratio in stars. 
In this case, however, we make use of the stellar mass distribution as a function
of O/Fe. Therefore we obtain:
\begin{equation}
<O/Fe>={1\over S_f} \int_0^{S_f} (O/Fe)(S)\,  dS\, ,
\label{PP75}
\end{equation}
where now $(O/Fe)(S)$ in the abundance ratio characterising the
gas out of which a mass $dS$ of stars formed.
This procedure will be repeated at each grid-point unless
specified otherwise.}

Then, we derive $[<O/Fe>]=log(<O/Fe>)-log(O/Fe)_{\odot}$, taking
the logarithm after the average evaluation (see Gibson, 1996).
Similar equations hold for [$<Fe/H>$] and the global metallicity [$<Z/H>$].

Another way to estimate the average composition of a CSP which is closer
to the actual observational value is to use the V-luminosity weighted abundances (which will be denoted
as $\rm <O/Fe>_V$). Following Arimoto \& Yoshii (1987), we have:
\begin{equation}
\rm <O/Fe>_V=\sum_{k,l}n_{k,l} (O/Fe)_l L_{V,k} / \sum_{k,l}n_{k,l} L_{V,k}    \, ,
\label{AY87}
\end{equation}  
where $n_{k,l}$ is the number of stars binned in the interval
centered around $\rm (O/Fe)_l$ with V-band luminosity $\rm L_{V,k}$.
Generally the mass averaged {\rm [Fe/H] and [Z/H]} are slightly larger than the luminosity
averaged ones, except for large galaxies (see Yoshii \& Arimoto, 1987, Matteucci et al., 1998). 
However there might be differences between the two methods at large radii,
as far as [Fe/H] and [Z/H] are concerned. In fact, the preliminary
analysis of PMC06 showed that both distributions may be broad and asymmetric 
and their mean values can provide a
poor estimate of the metallicity in complex systems with
a chemical evolution history quite extended in time. On the other hand,
PMC06 found the [Mg/Fe] distribution to be much more symmetric
and narrow than the [Z/H] distribution. Therefore, we expect that 
$[<O/Fe>] \simeq [<O/Fe>_V]$
at any radius and hence, we present mass-weighted values which are
more representative of the physical processes acting inside
the galaxy.
After PMC06, we will present our results in terms of $[<Fe/H>_V]$ and $[<Z/H>_V]$,
because the luminosity-weighted mean is much closer to the actual
observations and might differ from the average on the mass, unless otherwise stated.

Finally, in order to convert the predicted abundances for a CSP into
indices (especially in the case of short burst of SF), it is typically
assumed that a SSP with a \emph{mean} metallicity is representative of
the whole galaxy.  In other words, we use the predicted abundance ratios 
in stars for our CSPs to derive
the line-strenght indices for our model galaxies by selecting a SSP
with the same values for $[<O/Fe>],\, [<Fe/H>_V]\,$ and $\, [<Z/H>_V]$  
from the compilation of Thomas, Maraston \&
Bender (2003, TMB03 hereafter).

%{ In order to allow a meaningful comparison with the observation, all the 
%average abundance ratios are then projected along the line of sight.}

\subsection{Model description}

The present work is aimed at understanding the origin of the radial gradients 
in the stars by means of models which have photochemical properties
as well as radii comparable with those of typical massive ellipticals.
Moreover, we would like to understand what causes the $[<\alpha/Fe>]$ 
gradient slope to span the range
of values $\sim -0.2 - +0.2$ dex per decade in radius. 
In order to do that, we will essentially vary the initial conditions by adopting
reasonable hypotheses for the gas properties. 
A first classification of our set of models can be done according to their 
initial conditions (DM halo mass and available reservoir of gas): 

\begin{itemize}
%\item Model S: a $5.7\cdot 10^{12} M_{\odot}$ DM halo and $6.4\cdot 10^{11} M_{\odot}$ of gas   
\item Model M: a $2.2\cdot 10^{12} M_{\odot}$ DM halo and $\sim 2 \cdot 10^{11} M_{\odot}$ of gas
\item Model L: a $5.7\cdot 10^{12} M_{\odot}$ DM halo and $\sim 6.4\cdot 10^{11} M_{\odot}$ of gas
%\item Model cD: a $5.7\cdot 10^{13} M_{\odot}$ DM halo and $\sim 10^{13} M_{\odot}$ of gas
\end{itemize}
These quantitites have been choosen in order to ensure a final ratio between the mass of baryons
in stars and the mass of the DM halo around 0.1. Models by Matteucci (1992) and PM04 require
such a ratio for ellipticals in order to develop a galactic wind.
A more refined treatment of the link between baryons and DM
is beyond the scope of this work, and a more robust study of the gradient
creation in a cosmological motivated framework will be the topic of a forthcoming paper. 
The exact initial gas mass depends on the initial conditions 
and it is clear that gas can be accreted by the external environment. 
In particular, for each model we considered the following cases for the initial
gas distribution:

\begin{itemize}
\item[a:] isothermal density profile. { In this case, the gas is assumed
to start from an isothermal configuration of equilibrium within the galactic (i.e. considering both
DM and gas) potential well. The actual initial temperature is lower than the virial
temperature, in order to induce the gas to collapse.
These initial conditions might not be justified by the current Cold DM paradigm for
the formation of structures. However, we consider them very useful because
they give the closest approximation of the typical initial conditions
adopted by the chemical evolution models to which we will compare our results.
The reader can visualise this model as an extreme case in which we let all the gas
be accreted before the SF starts}
\item[b:] { constant density profile. In this case the gas has an initial value for the mass
density which is constant with radius in the whole computational box (c.f.sec 2.1). 
The DM and, afterwards, the gas and stellar gravity 
will then create the conditions for a radial inflow to happen.This case might be more realistic
than the former one, in the sense that the DM potential will ``perturb'' the gas which
is uniformely distributed at the beginning of the simulation.
At variance with the previous model, in this case we let the SF
process start at the same time at which the gas accretion starts.}
%\item[c:] both DM and gas haloes accreted during the evolution
\end{itemize}

%Unless otherwise stated, the gas has an intial temperature set by the virial theorem.
Table 1 summarises the main properties of each model that will be discussed in this paper, namely
the core radius for both the DM and the gas profile, the SF parameter $\epsilon_{SF}$,
the initial temperature and  the SN efficiency $\epsilon_{SN}$ respectively.

Concerning the class of models labelled \emph{a}, we mainly vary the gas temperature and the 
parameter of star formation. We do not vary the gas mass (via the core gas density {
and radius}) because we need that precise amount of gas in order to ensure that: i) enough 
stars can be created; ii) 
at the same time there is not too much gas left (we recall that present-day ellipticals are basically without gas).
{ Also, the assumed profile guarantees the most of the gas is already within the final
effective radius of the galaxy in a way which mimick the assumptions made in PM04 and PMC06.}

For the class of models labelled \emph{b}, instead, the initial gas density (as
reported in Table 1 under the column pertaining $\rho_{core,gas}$ ) can be a 
crucial parameter, as well as the gas temperature and $\epsilon_{SF}$.
Here the values for $\rho_(r,t=0)=\rho_{core,gas}$  
%may be lower than in the \emph{a} cases,
%otherwise we would have too much gas in the grid, even 
{ is chosen in order to have the initial gas content in the whole grid not} higher than the typical
baryon fraction in high density environment (i.e. 1/5-1/10 as in 
galaxy cluster, e.g. McCarthy et al. 2007).
In each case, the gas temperature ranges from $10^{4-5}$ K (cold-warm gas) 
to $10^{6-7}$ K (virialised haloes).
We limit both the DM and the stellar profile to their tidal radii,
chosen to be 66 kpc (both of them) in case M as well as 200 kpc and 100 kpc, 
respectively,
in case L. These values are consistent with the radii of the X-ray haloes 
surrounding ellipticals of the same mass. 

%\par
%\hbox{}
\begin{table*}
\centering
\begin{minipage}{120mm}
\scriptsize
\begin{flushleft}
\caption[]{Input parameters}
\begin{tabular}{l|llllllll}
\hline
\hline
Model	&$R_{core,DM}$ & $R_{core,gas}$&  $\rho_{core,gas}$ &$\epsilon_{SF}$ 	&  T   & $\epsilon_{SN}$ \\
        & ({kpc})      & ({kpc})      &  ($10^{-25}\rm g\, cm^{-3}$)     & &   (K) &    	     \\
\hline

%Sa  (jnew)  &0.1   &  0.1           &                      &0.5              &$10^5$	& 0.1 	\\
%Sb1 (jbis)  &0.1   &  -             &                      &0.5              &$10^5$	& 0.1 	\\
%Sb2 (jter)  &0.1   &  -             &                      &3              &$10^5$	& 0.1 	\\
%\hline
%\hline

Ma1  & 1.5  &  0.4         &   0.6                &1              &$10^6$	& 0.1\\
Ma2  & 1.5   &  0.4           &   0.6             &10              &$10^4$	& 0.1\\
Ma3  & 1.5   &  0.4           &   0.6               &2              &$10^4$	& 0.1\\
MaSN  & 1.5   &  0.4           &   0.6              &1              &$10^6$	& 1.0\\
Mb1   & 1.5   &  -            &  0.06               &1              &$10^7$	& 0.1 	\\
Mb2  & 1.5   &  -            &  0.2                &1              &$10^5$	& 0.1	\\
Mb3   & 1.5   &  -            &  0.06                &10              &$10^6$	& 0.1	\\
Mb4   & 1.5   &  -            &  0.6               &1              &$10^6$	& 0.1 	\\
Mb5  & 1.5  &  0.4           &   0.6               &2              &$10^4$	& 0.1\\
%Mc (H)       & -        &  -         &  ?                  &10              &$10^4$	& 0.1\\
\hline

La   & 4.5   &  1.0           &   0.6              &10              &$10^7$	& 0.1	\\
Lb & 4.5   &  -           &   0.6             &10              &$10^6$	& 0.1	\\

%\hline
%\hline
%cDa1 (p)   & 	\\
%cDa2 ()  & 	\\
%cDb (pqtr)& 	\\

\hline
\end{tabular}
\end{flushleft}
\end{minipage}
\end{table*}

\section{Results: a general overview}

%\subsection{A general overview}

The main results of our models are presented in Table 2, where the
final (i.e. after SF stops) values for the stellar core and
effective radii, the time for the onset of the galactic wind in the central
region ($t_{gw}$), the abundance ratios in the galactic center and the
gradients in [$<O/Fe>$] and [$<Fe/H>_V$], are reported.  In
particular, we choose $R_{eff,*}$ as the radius which contains 1/2 of
the stellar mass of the galaxy and, therefore, it is directly
comparable with the observed effective radius, whereas $R_{core,*}$ is
the radius encompassing 1/10 of the galactic stellar mass. In most
cases, this radius will correspond to $\sim 0.05 - 0.2 R_{eff,*}$,
which is the typical size of the aperture used in many observational
works to measure the abundances in the innermost part of ellipticals.
We did not fix $R_{core,*} = 0.1 R_{eff,*}$ a priori, in order to have
a more meaningful quantity, which may carry information on the actual
simulated stellar profile.  Finally, we did use the following
notation for the metallicity gradients in stars
$\Delta_{O/Fe}=([<O/Fe>]_{core}-[<O/Fe>]_{eff})/log
(R_{core,*}/R_{eff,*})$; a similar expression applies for both the
[$<Fe/H>_V$] and the [$<Z/H>_V$] ratios.

The slope is calculated by a linear regression between the core and
the half-mass radius, unless otherwise stated.  Clearly, deviations
from linearity can affect the actual slope at intermediate radii.
Before discussing in detail the galactic formation mechanism of our models, we
must check whether they resemble typical ellipticals for a given mass.
First of all, we have to ensure that the MMR is satisfied.  The
majority of our model galaxies exhibits a central mean values of
[$<Fe/H>_V$] within the range inferred from integrated spectra, namely
from -0.8 to 0.3 dex (Kobayashi \& Arimoto 1999). On average, the more
massive galaxies have a higher metal content than the lower mass ones. 
However, the small range in the final
stellar masses as well as the limited number of cases
presented here prevent us from considering our models as a 
complete subsample of typical ellipticals drawn
according to some galactic mass function.
Here we simply check whether our models fullfill the
constraints set the MMR and the CMR for a galaxy of $\sim 10^{11} M_{\odot}$.

For instance, we applied the Jimenez et al. (1998) photometric code to
both cases Ma1 and La (inside their effective radius), and found the
results in good agreement with the classic Bower et al. (1992) CMRs.
In fact, by assuming an age of 12.3 Gyr (which in a standard
Lambda CDM cosmology means a formation redshift of 5), we have $M_V=-20$ mag,
U-V=1.35 mag, V-K=2.94 mag and J-K=0.97 mag for model Ma1, whereas for
the case La we predict $M_V=-21.3$ mag, U-V=1.28 mag, V-K=3.17 mag
and J-K=1.06 mag.  It can be shown that similar results apply to all
the other cases, because their star formation histories as well as
their mean metallicity are roughly the same. It is known, in fact,
that broad-band colours can hardly discriminate the details of a SF
episode if this burst occurred long ago in the past.

The models show an average [$<\alpha/Fe>$] = 0.2 - 0.3 as requested by 
the observations 
(Worthey et al. 1992, Thomas et al. 2002, Nelan et al. 2005).
In general, the predicted abundance ratios are consistent with the reported 
$\sim 0.1$ dex-wide observational scatter of the above mentioned articles, with
the exception of a few cases which will be discussed in the following sections.

On the other hand, several models (not presented here) matching the chemical properties
fail in fitting other observational constraints.
As an example, here we report model Mb5, whose stellar core radius is by far
too large to be taken into account in the remainder of the paper.

Model MaSN, instead,  shows how a strong feedback from SN
can suppress the SF process too early, as testified by the high  
predicted $\alpha$-enhancement in the
galactic core. Also in this case the galaxy
is too diffuse. 
It can be shown that $\epsilon_{SN}$ in the range 0.1-0.2 does 
not lead to strong
variations in the results. Therefore, we adopt $\epsilon_{SN}$= 0.1, in line
with the calculations by Thornton et al. (1998).

In all the other cases, the dimension of the model galaxies (i.e. their
effective radii) are consistent with the values reported for bright ellipticals
(e.g. Graham et al. 1996).

We stress that here we are not interested in a further fine tuning of
the input parameters in order to reproduce the \emph{typical average
elliptical} as in PM04.  Our aim is, instead, to understand whether it is
possible to explain the observed variety of [$<O/Fe>$] gradient slopes
\emph{once} all the above constraints have been satisfied.  In order to do
this we first examine the formation of the stellar component of a
typical elliptical galaxy.  Then we derive further constraints by comparing
both the predicted abundance and line-strength indices gradients with
observations. Finally, we study in great detail the role of several
factors in shaping the [$<O/Fe>$] gradients.

%\par
%\hbox{}
\begin{table*}
\centering
\begin{minipage}{120mm}
\scriptsize
\begin{flushleft}
\caption[]{Model results}
\begin{tabular}{l|llllllll}
\hline
\hline
Model	&$M_*$         &$R_{core,*}$ & $R_{eff,*}$& $t_{gw,core}$ & [$<O/Fe>_{*,core}$] & [$<Fe/H>_{*,core}$] & $\Delta_{O/Fe}$ &  $\Delta_{Fe/H}$   \\
        &($10^{10}M_{\odot}$) &({kpc})      & ({kpc})   &  (Myr)        &                     &                     &                 &                     \\
\hline

%Sa  (jnew)  & 0.74   &0.1     & 1.7	& ???            &  0.08               & -0.35                 &  -0.04          & -0.26		\\
%Sb1 (jbis) & 0.74   & 0.1     & 1.7	& ???            &  0.36               & -0.50                 &  -0.13          & -0.21	\\
%Sb2 (jter)  & 0.74 & 0.1     & 1.7	& ???            &  0.28               & -0.07                 &  -0.11          & -0.21	\\
%\hline
%\hline

Ma1 &6.0&  0.3     & 12	& 1100             &  0.29               & 0.13                 &  0.02          & -0.13	\\
Ma2    &25.&  0.4     & 7.7	& 800             &  0.22               & 0.35                 &  -0.21          & -0.16	\\
Ma3   &25.&  0.4     & 8.3& 800             &  0.35               & 0.57                &  -0.17         & -0.03	\\
MaSN    &2.0 & 6.6    &  31 & 200              &  0.55               & -0.51                 & -0.14            &  +0.27      \\
Mb1   &6.0 & 	0.4     & 17	& 700             &  0.14               &  0.22                &  0.09          & -0.31	\\
Mb2   &3.0 & 0.2     & 8.7	& 300             &  0.33               & -0.02               &  0.          & -0.18	\\
Mb3   &21&  0.4     & 8.8	& 440             &  0.17            & 0.37               &  -0.08        &  -0.29	\\
%Mb3 (cter)  &6.4&  11    & 33	& 500             &  0.36            & 0.45              &  -0.08        &  -0.40	\\
Mb4  &26 & 	0.4     & 5.4	& 200             &  0.42               &  -0.40                &  -0.08          & -0.20	\\
Mb5  &25.&  14.8     & 33.6& 1400             &  0.36               & 0.17                &  -0.14         & -1.40	\\
%Mc (H)    &2.0 & ?    &  23 & ---              &  ???               & ???                 & ???            &        \\
\hline

La    &26& 3.4     & 29	& 400             &  0.14               & 0.70                 &  0.19           & -0.50\\
Lb   &29&  2     & 21	& 400             &  0.12               & 0.57                 &  0.32           & -0.50	\\

%\hline
%\hline
%cDa1 (p)   & 	\\
%cDa2 ()  & 	\\
%cDb ()& 	\\

\hline
\end{tabular}
\end{flushleft}
Values predicted after the SF has finished.
\end{minipage}
\end{table*}

\subsection{The outside-in formation of a typical elliptical}

\subsubsection{The gas-dynamical evolution}

\begin{figure}
%\epsscale{.80}
\includegraphics[width=8cm,height=8cm]{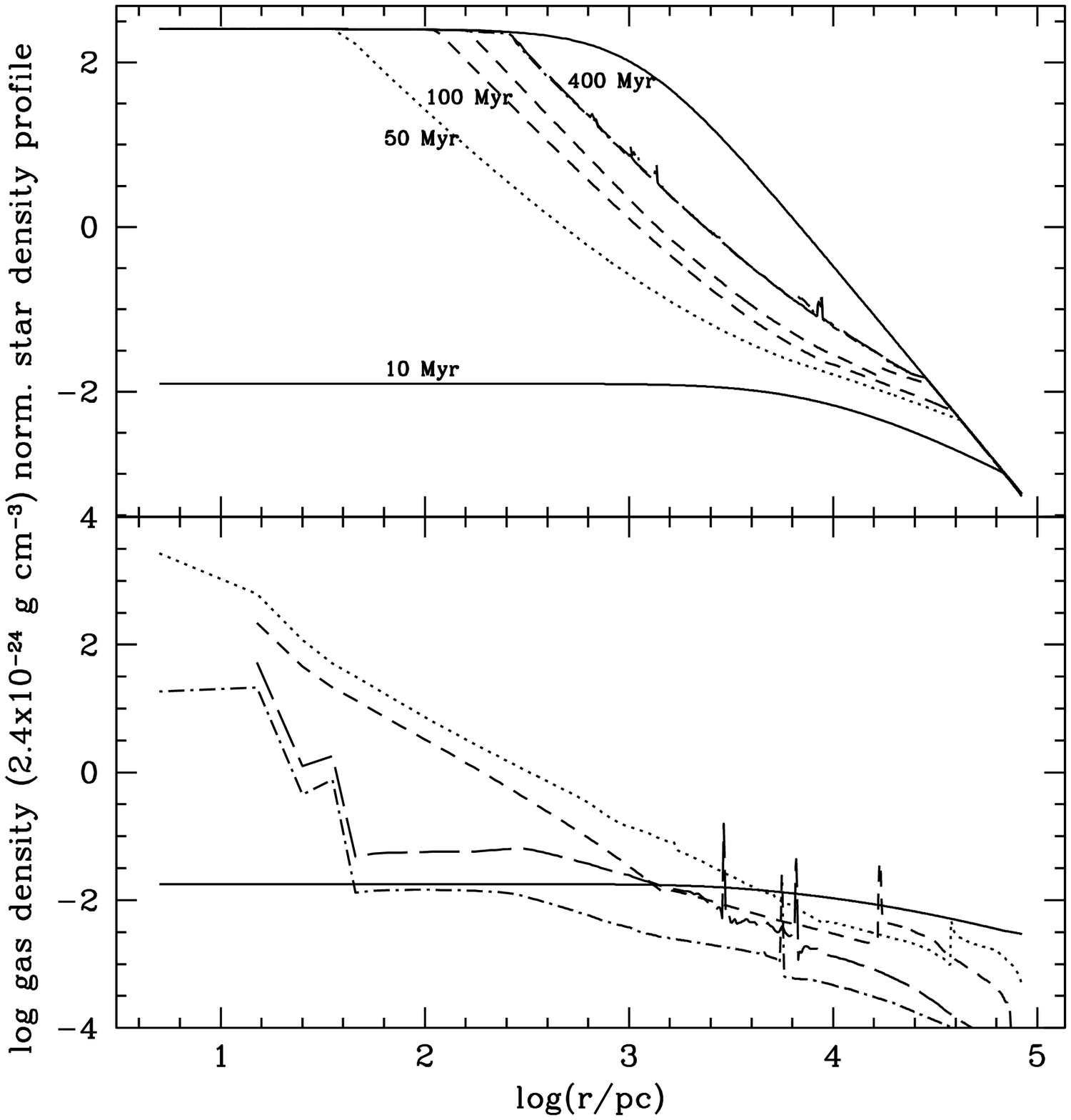}
\includegraphics[width=8cm,height=8cm]{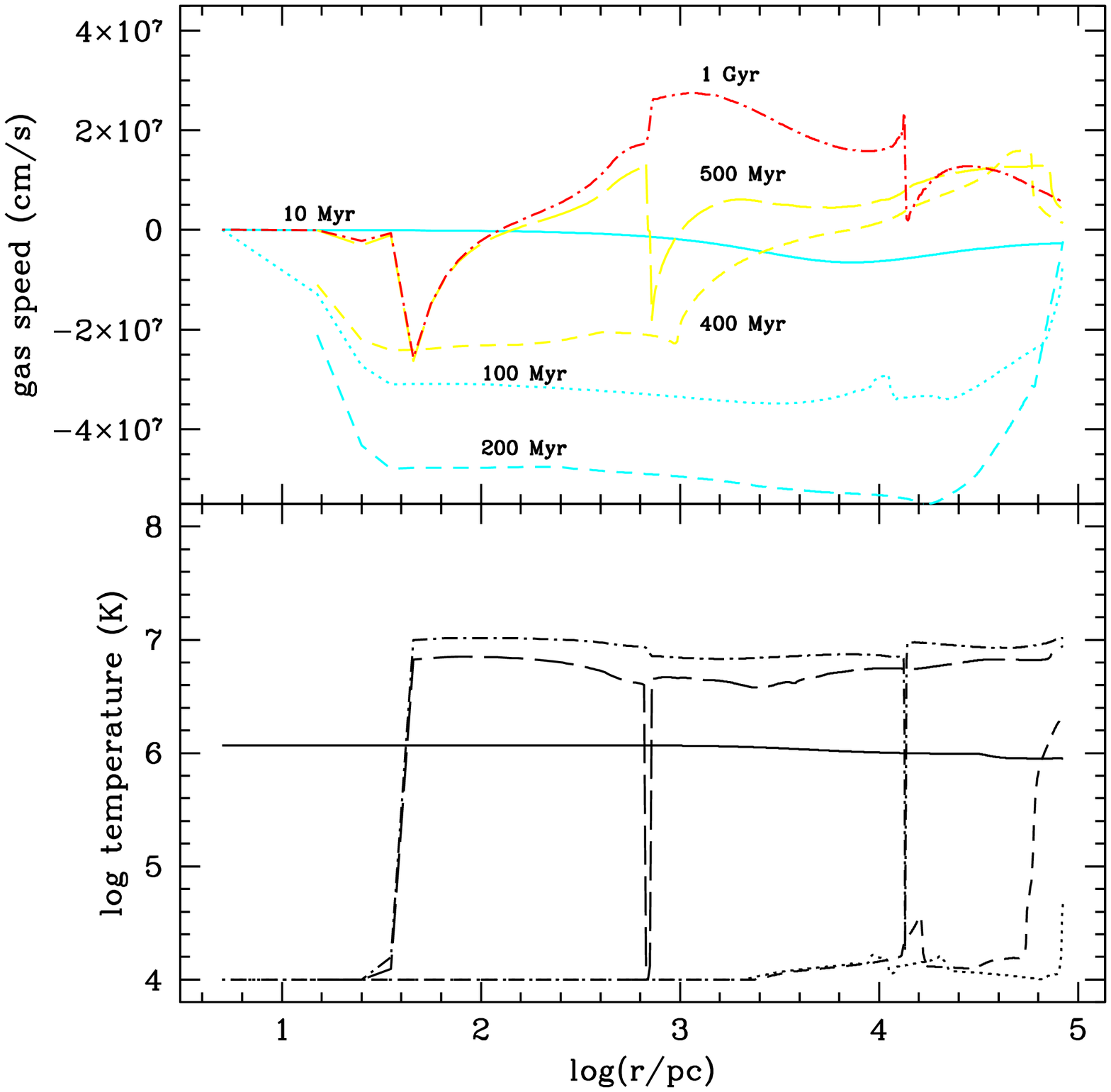}
%\plotone{fig2a.eps}
\caption{\emph{Upper panels}: the stellar mass- (top) and gas density
(bottom) profiles predicted by model La at different times: 10 Myr
(solid), 50 Myr (dotted), 100 Myr and 200 Myr(dashed), 400 Myr
(dotted-dashed). The model predictions at 1 Gyr coincide with the ones
at 400 Myr. The thick solid line without time labels represents a King profile (see text).
\emph{Lower panels}: the gas velocity (top) and temperature (bottom)
profiles predicted by model La at different times: 10 Myr (solid),
100 Myr (dotted), 200 Myr (thick-dashed), 400 Myr and 500 Myr
(thin-dashed), 1 Gyr (dotted-dashed).}
\label{hydro_q}
\end{figure}

\begin{figure}
%\epsscale{.80}
\includegraphics[width=8cm,height=8cm]{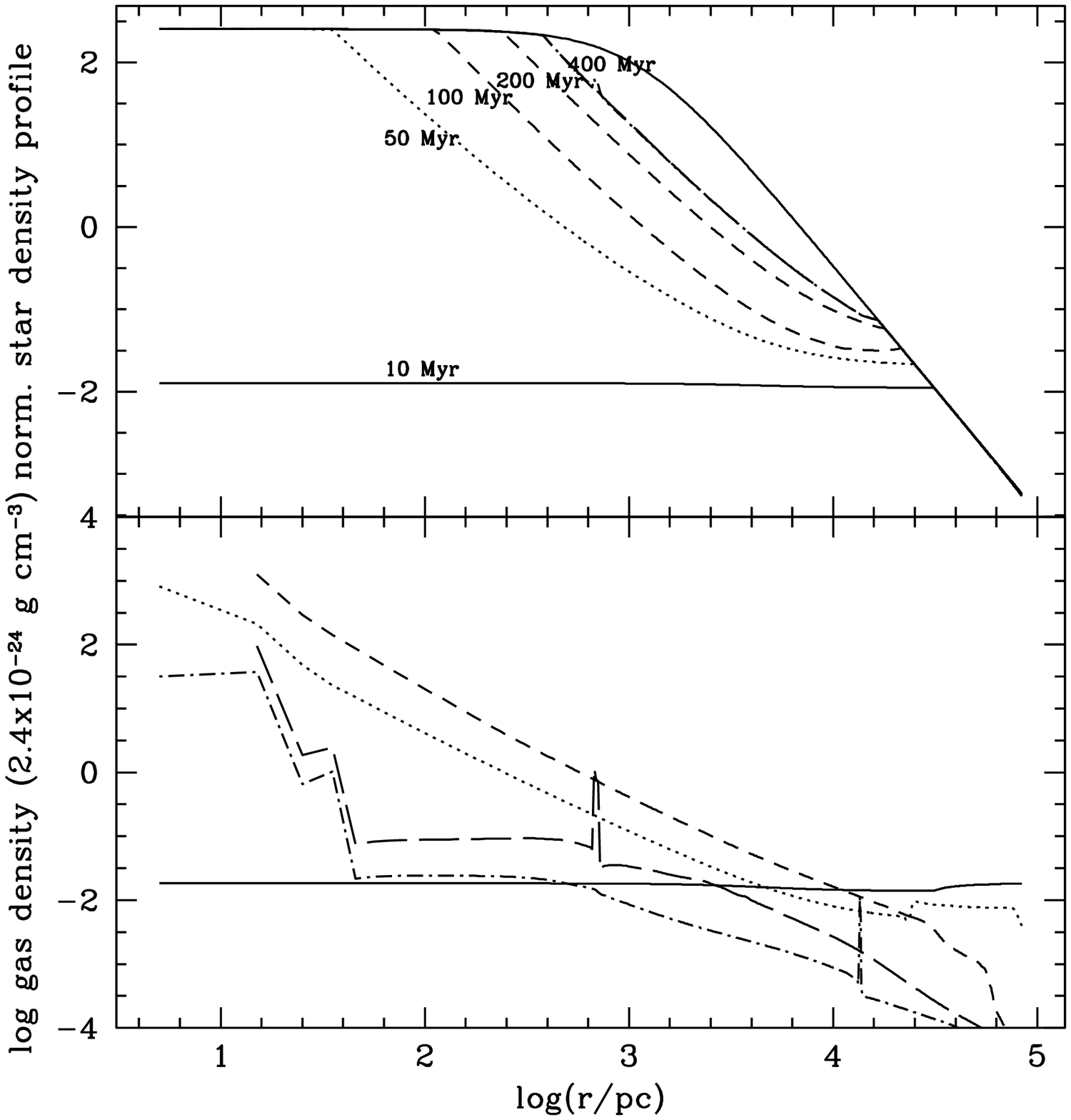}
\includegraphics[width=8cm,height=8cm]{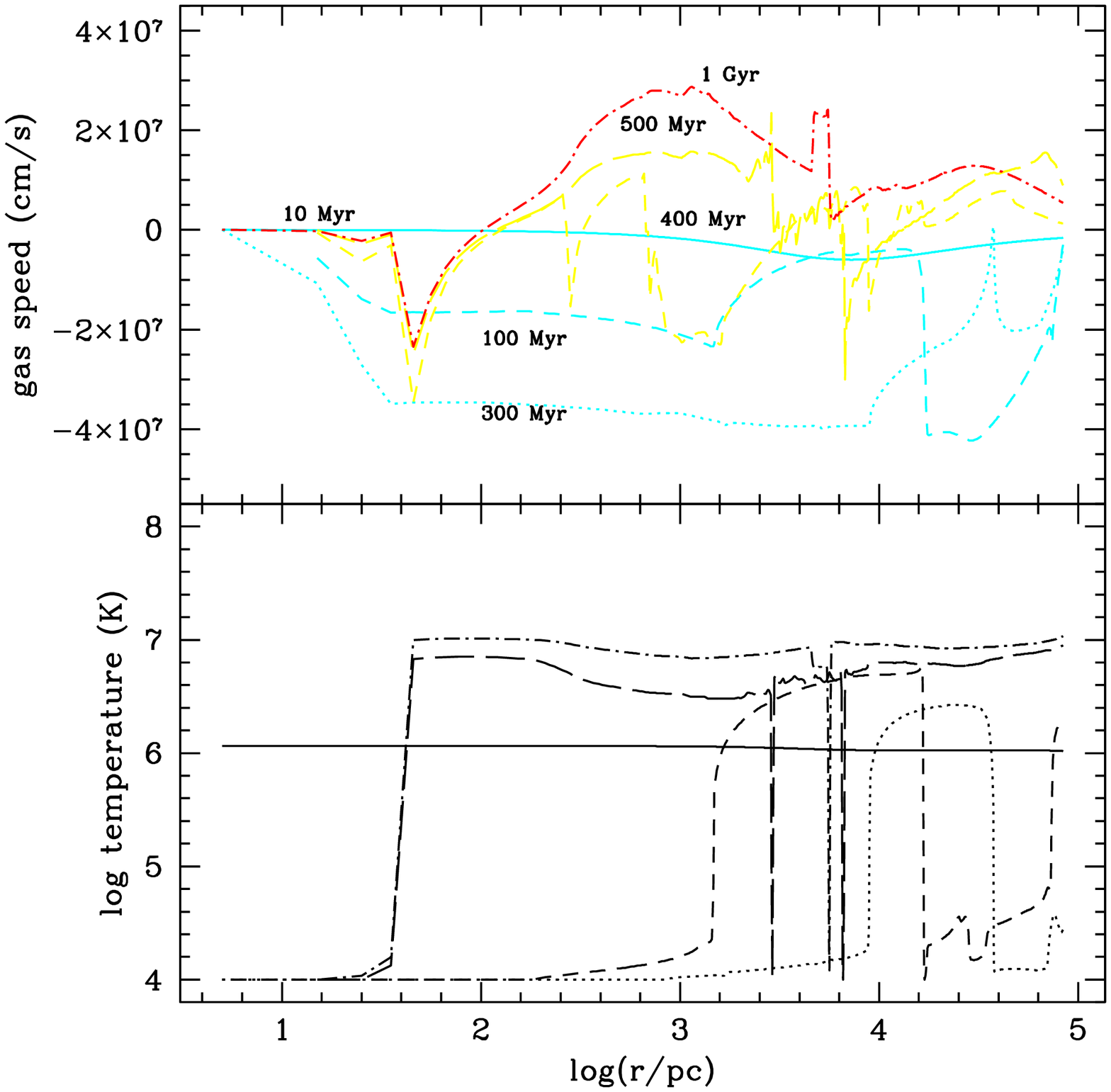}
\caption{Temporal evolution of density, velocity and temperature profiles for model Lb.
The meaning of the curves is the same of Fig.~\ref{hydro_q}}
\label{hydro_qqtr}
\end{figure}

In this section we focus on the formation mechanism of a single galaxy:
the time evolution of its abundance gradients will be the subject of Sec. 4.1 .  
A clear example of a
massive elliptical is given by model La (massive elliptical with
the gas in intial equilibrium at $10^7$ K and $\epsilon_{SF}$=10), whose chemo-dynamical
evolution is shown in  Figs.~\ref{hydro_q},\ref{fet}-\ref{grad_evol}.  
We will refer to this particular model as a reference case for characterizing
the hydrodynamical behaviour of our models, as well as to derive general hints
on both the development of the metallicity gradients and the SF process.
We will also compare the results of models La with those of models Lb, 
being the main difference between the two models in the 
initial gas distribution. 
Fig.~\ref{hydro_q} shows the stellar and the gas density
profiles (upper panels) as well as the gas velocity and the
temperature profiles (lower panels) at different times (see captions
and labels).  It can be clearly seen that at times earlier than 300
Myr the gas is still accumulating in the central regions where the
density increases by several orders of magnitude, with a uniform speed
across the galaxy. The temperature drops due to cooling, and the SF
can proceed at a very high rate ($\sim 10^{2-3} \rm M_{\odot}
yr^{-1}$).  In the first 100 Myr the outermost regions are built-up,
whereas the galaxy is still forming stars inside its effective
radius. For comparison, the thick solid line in the star density panel
shows the adopted threshold (King profile).  We show the
evolution predicted by model Lb (similar to La, but with an initial
accretion of gas) in Fig.~\ref{hydro_qqtr}. We notice, that, despite
the different initial conditions, the evolution of all the physically
interesting quantities follows the results obtained for model La.

After 400 Myr, the gas speed becomes positive (i.e. outflowing gas) at
large radii, and at a 500 Myr almost the entire galaxy is experiencing a
galactic wind. This model proves that a massive galaxy can undergo a
galactic wind, which develops outside-in, thanks to the sole energy
input from SNIa+II.  The wind is supersonic for, at least, the first
Gyr after $t_{gw}$,which is the time of the onset of the galactic
wind and depends on the model assumptions.  At roughly 1.2 Gyr, 
the amount of gas left inside the galaxy is below 2\% of the stellar mass.  
This gas is
really hot (around 1 keV) and still flowing outside.  Therefore, as
anticipated also by our chemical evolution studies (Pipino et
al. 2002, PM04, Pipino et al. 2005), a model with Salpeter IMF and a value for $\epsilon_{SN}=0.1$ can mantain a strong galactic wind for
several Gyr, thus contributing to the ejection of the chemical
elements into the surrounding medium.

{ The fact that the galactic wind occurs before externally than internally 
is simply due to the fact that the work to extract the gas from the outskirts is smaller 
than the work to extract the gas from the center of the galaxy. Therefore, since the galactic 
wind occurs first in the outer regions the star formation rate stops first in these regions, 
for lack of gas. In the following we will refer to \emph{the outside-in scenario} as to the fact that the SFR halts before 
outside than inside due to the progressive occurrence of the galactic wind from outside to inside.}

\subsubsection{Chemical abundances: from the gas to the stars}

\begin{figure}
%\epsscale{.80}
%\plottwo{fig3a.eps}{fig3b.eps}
%\plotone{fet_bbis.eps}
\includegraphics[width=8cm,height=8cm]{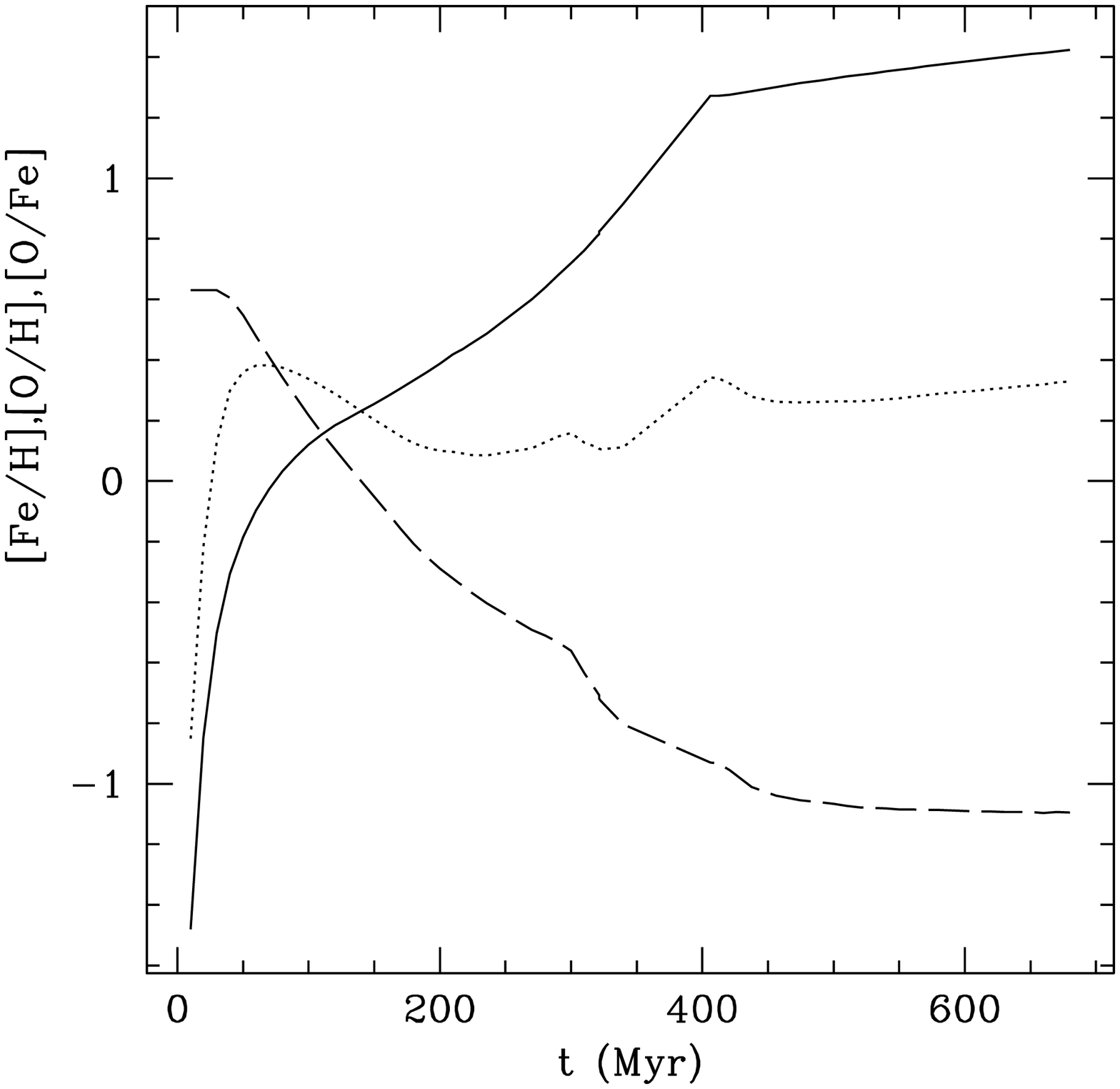}
\caption{Time evolution of [Fe/H] (solid), [O/H] (dotted), [O/Fe] (dashed) in the gas of model La.
These abundances are values for the whole galaxy.}
\label{fet}
\end{figure}

\begin{figure}
%\epsscale{.80}
\includegraphics[width=8cm,height=8cm]{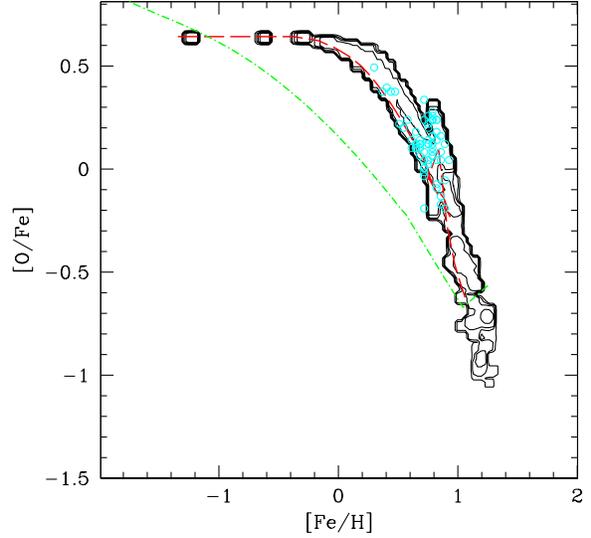} %{cont_new_core.eps}
\includegraphics[width=8cm,height=8cm]{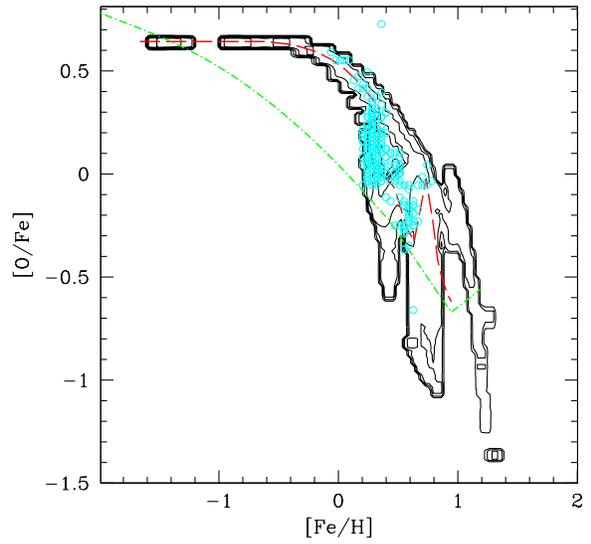} %{cont_new_reff.eps}
\caption{Contours: bidimensional metallicity distribution of stars as
functions of [Fe/H] and [O/Fe] for the core (upper panel) and
the effective radius regions (lower panel) of model La. 
Dots: ramdomly generated stars in order to emphasize the peaks in the distributions.
Dashed line: [O/Fe] vs. [Fe/H] in the gas of model La (mass-weighted values
on the gridpoints of each region).
Dot-dashed line: [O/Fe] vs. [Fe/H] in the gas, as predicted by the best model of PM04
for a galaxy with similar stellar mass.}
\label{gdwarf}
\end{figure}

\begin{figure}
%\epsscale{.80}
\includegraphics[width=8cm,height=8cm]{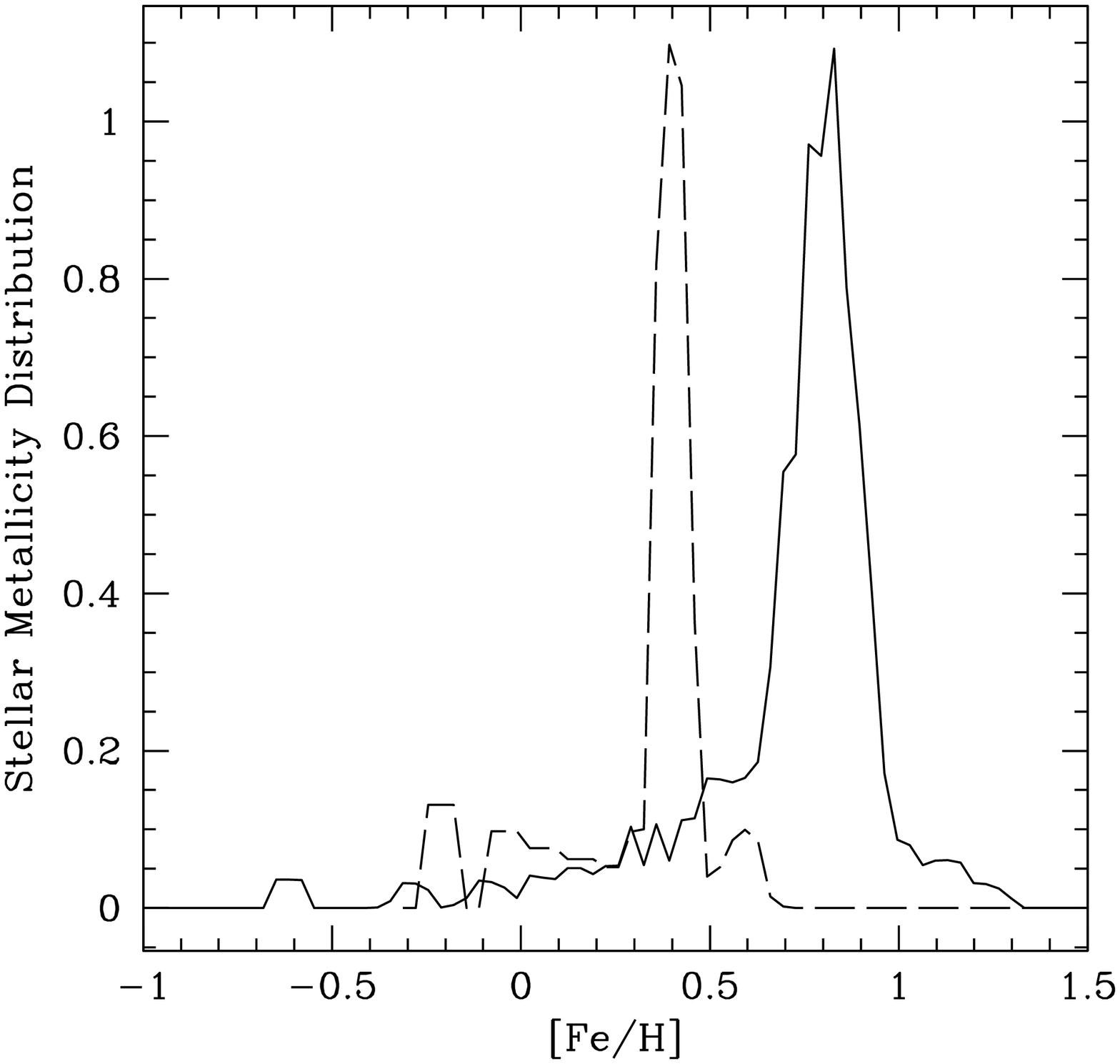}
\caption{The final Stellar Metallicity Distribution as a function of [Fe/H]
for model La. The values have been arbitrarily rescaled. The two
peaks represent the different chemical enrichment
suffered at different radii (see text). The solid
line refers to the galactic core radius, whereas the dashed line
is the prediction for a shell 5 kpc wide, centered at $R_{eff}$.}
\label{gdwarf_1}
\end{figure}

%In order to understand the values in the mean stellar abundances, we
%need to make a step backward and analyse the abundance ratios in the
%gas.  

In Fig.~\ref{fet} we show the temporal evolution of the
elemental abundances in the gas for the entire galactic volume. 
As expected, the prompt release of O
by SNII makes the [O/H] in the gas to rise very quickly, whereas the Fe
enrichment is delayed.  As a result, the [O/Fe] ratio spans nearly two
orders of magnitude, reaching the typical value set by the SNIa yields
after 500 Myr.  We can derive much more information from
Figs.~\ref{gdwarf}, where the metallicity distribution
of stars as a function of [$Fe/H$] and [$O/Fe$] are shown.  In these
figures we plot the distribution of stars formed out of gas with a
given chemical pattern (i.e. a given [$Fe/H$] and [$O/Fe$]) as
contours in the [$O/Fe$]-[$Fe/H$] plane. In particular, the contours
connect regions of the plane with the same mass fraction of stars.  
Since we consider the stars
born in different points of the grid, which may have undergone
different chemical evolution histories, it is useful to focus on two
different regions: one limited to $R_{core,*}$ (upper panel) and the
other extending to $R_{eff,*}$ (lower panel).  It is reassuring that
in both panels the overall trend of the [$O/Fe$] versus [$Fe/H$] in the
stars agrees with the theoretical plot of [O/Fe] versus [Fe/H] in the
gas expected from the time-delay model (Matteucci \& Greggio
1986).  For comparison, 
we plot the output of PM04's best model with roughly the same stellar
mass as a dot-dashed line in fig.~\ref{gdwarf}.
Both the early and final stages of the evolution coincide.
An obvious difference is that the \emph{knee} in the [O/Fe] vs [Fe/H] 
relation predicted by our model is much more evident than the one of PM04.
The reason must be ascribed to the fact that here we adopt a fixed
O/Fe ratio in the ejecta of SNII, whereas the stellar yields show
that there is a small dependence on the progenitor mass (which is taken into
account in detailed chemical evolution models as the PM04 one).
Moreover, as we will show in Sec. 5.1, most of the metals locked-up in the
stars of the galactic core were produced outside the core.
In practice, we anticipate that
the inner regions suffer a metal-rich initial infall (i.e. inflowing gas
has a higher [Fe/H] abundance with respect to the gas already present
and processed in the inner regions),
therefore the number of stars formed at $[Fe/H]\le -1$ is very small
compared to number of stars created at very high metallicities.
This fast increase of the [Fe/H] ratio in the gas also makes the 
\emph{knee} of the upper panel of Fig.~\ref{gdwarf} more evident than
the one in the lower panel.
\footnote{The physical mechanisms which produce such a metal-enhanced 
internal gas flows,
as well as their role in changing the [O/Fe] ratio in the gas, will 
be discussed in great detail in Sec. 5.}

The above results have two implications: first, the fact that our
implementation of the chemical elements in the hydrodynamical code does
not produce spurious chemical effects and it has been done in the proper
way.  Second, and perhaps more important, it shows that a chemical
evolution model gives accurate predictions on the behaviour of the mean
values, even though it does not include the treatment of gas radial
flows and it has a coarser spatial resolution.  As expected from the
preliminary analysis of PMC06, the innermost zone (Fig. 4, upper panel) 
exhibits less scatter.  At larger radii, the
distribution broadens and the asymmetry in the contours increases.  This can be more
clearly seen in the classical G-dwarf-like diagram of
Fig.~\ref{gdwarf_1}, where the number of stars per [Fe/H] bin only 
is shown. 
%From the comparison
%between these model predictions and the observed G-dwarf-like diagrams
%derived at different radii by Harris \& Harris (2002, see their
%fig. 18) for the elliptical galaxy NGC 5128, we can derive some
%general considerations.  The qualitative agreement is remarkable: we
We can explain the smooth early rise in the [Fe/H]-distribution in the inner 
part (solid
line) as the effect of
the initially infalling gas, whereas the sharp truncation at high
metallicities is the first direct evidence of a sudden and strong wind
which stopped the star formation.
The suggested outside-in formation
process reflects in a more asymmetric shape of the G-dwarf diagram at
larger radii (dashed line), where the galactic wind occurs earlier (i.e. closer to
the peak of the star formation rate), with respect to the galactic
centre.  The broadening of the curves, instead, reflects the fact that the outer
zone (extending to $R_{eff,*}$) encloses several shells with different
SF as well as gas dynamical histories.

In practice, the adopted [$<Fe/H>$] and [$<O/Fe>$] are either the
mass or the luminosity weighted values, taken from the distributions
similar to the one of fig.~\ref{gdwarf_1} (but in linear scale) 
according to eq.~\ref{PP75} and~\ref{AY87} .
They can be compared with SSP-equivalent values inferred from the
observed spectra taken from the integrated light (see next Section).  
These quantities tell us that, models La and Lb exhibit a quite 
high [$<Fe/H>_V$]
in the stars of the galactic core, although model Lb is in
slightly better agreement with the observed central values of [$<Fe/H>_V$] (Carollo et al. 1993,
Mehlert et al. 2003, Sanchez-Blazquez et al. 2006) than model La.
%A this stage, however, we are still interested in the showing , 
%therefore we keep on showing model La's evolution...irrespective of the zero point (i.e. the core
%values of the abundance ratios)...

\section{The formation of the abundance gradients}

\subsection{The temporal evolution of the gradients in the reference case}

\begin{figure}
%\epsscale{.80}
%\plottwo{fig3a.eps}{fig3b.eps}
%\plottwo{bbis_grad.eps}{bbis_grad_ofe.eps}
%\includegraphics[width=8cm,height=8cm]{q_grad.eps}
\includegraphics[width=8cm,height=8cm]{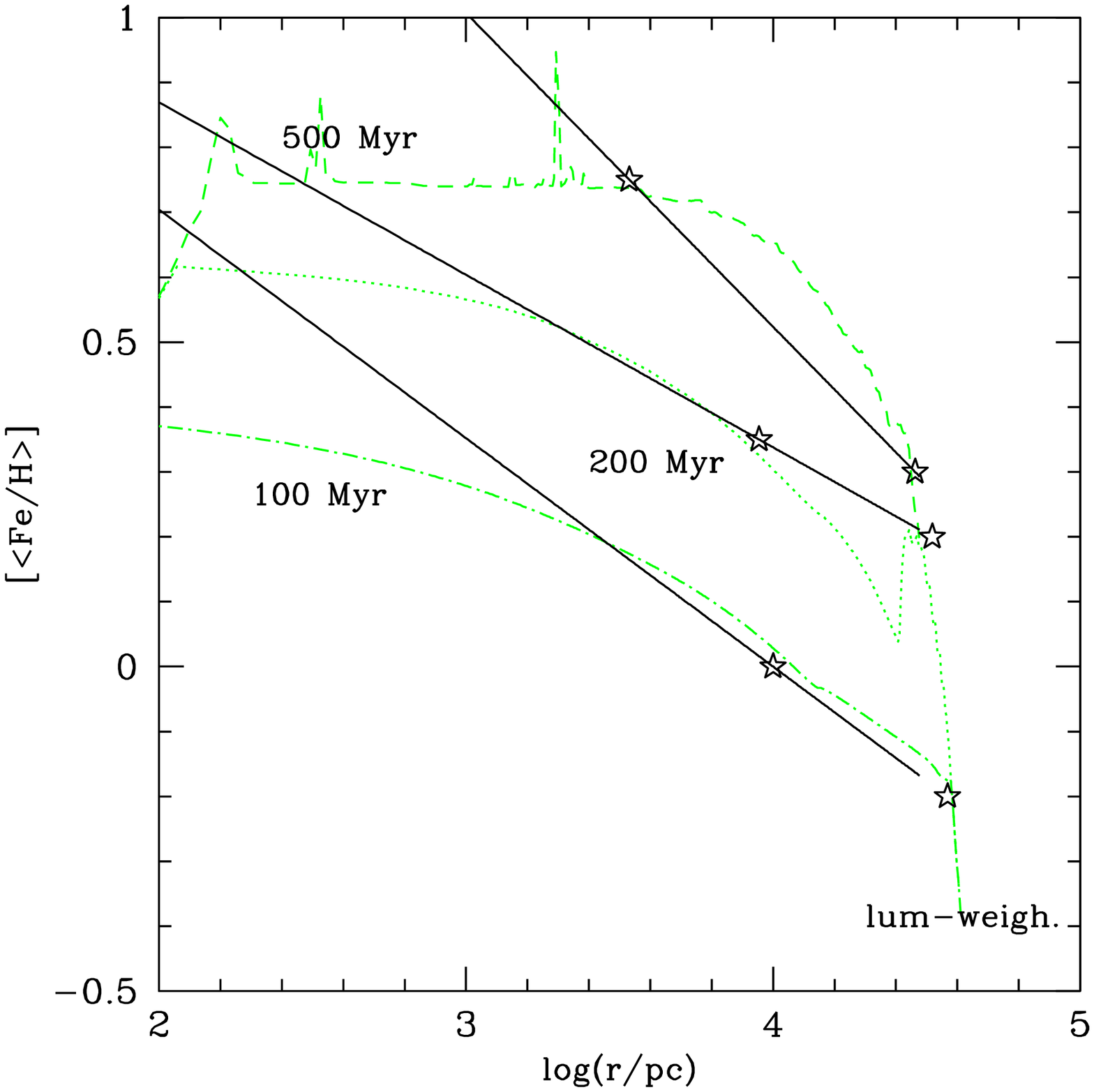}
\includegraphics[width=8cm,height=8cm]{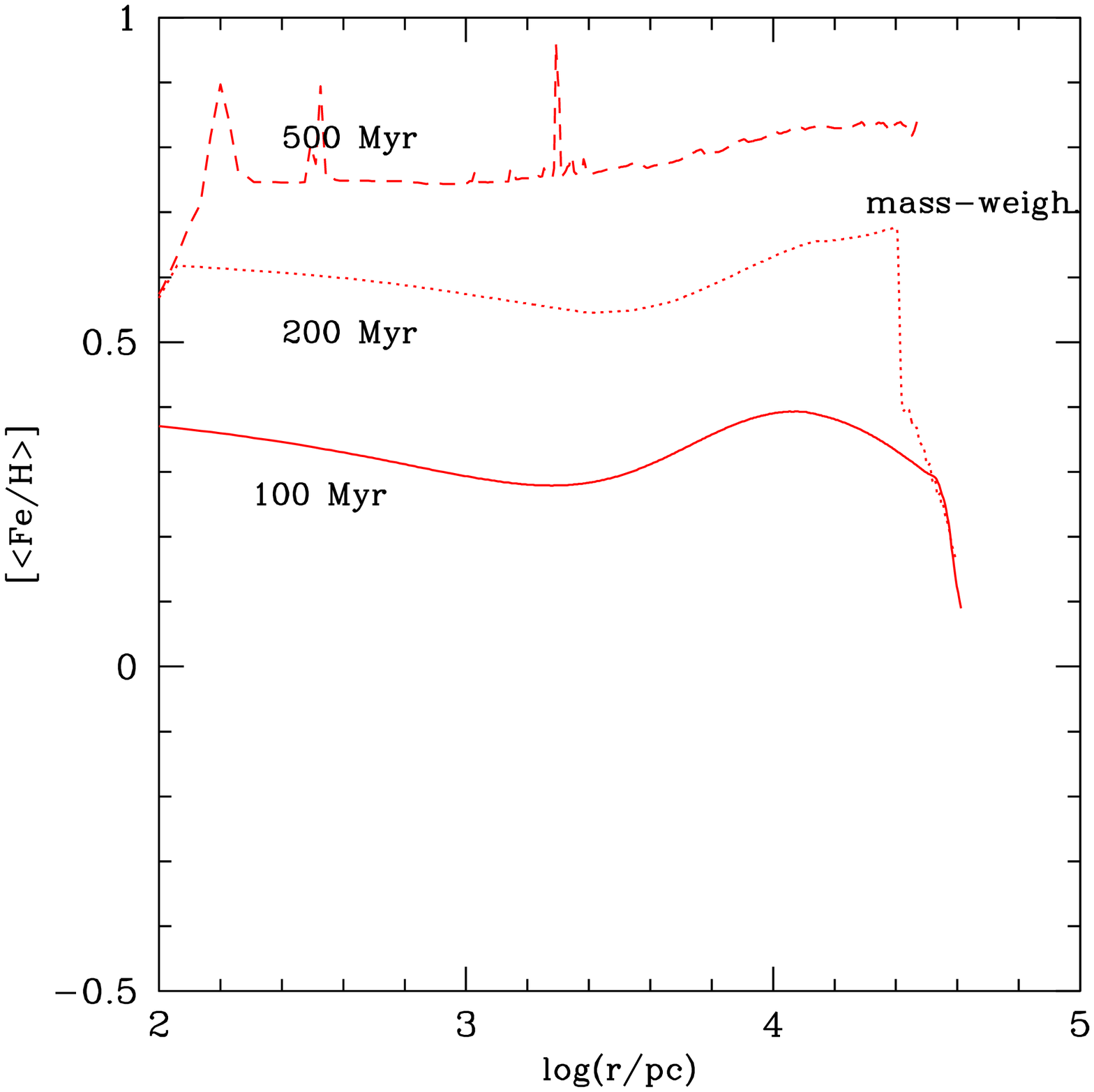}
\caption{Time evolution of radial metallicity gradients in stars predicted by model La. 
\emph{Upper panel}: the luminosity weighted [$<Fe/H>_V$] in stars versus 
radius at different times
(dotted-dashed: 100 Myr; dotted: 200 Myr; dashed: 500 Myr).
The stars mark the luminosity-weighted values at both the core and 
effective radius.
The solid lines represents the gradients inferred by a simple linear regression
fit of the values at both the core and effective radius at each time.
\emph{Lower panel}:
The mass-weighted [$<Fe/H>_*$] in stars versus radius at different times 
(as above).
The scale is the same of the upper panel. We remark the differences
between mass- and luminosity weighted quantities at large radii.}
\label{grad_evol}
\end{figure}

\begin{figure}
%\epsscale{.80}
%\plottwo{fig3a.eps}{fig3b.eps}
%\plottwo{bbis_grad.eps}{bbis_grad_ofe.eps}
\includegraphics[width=8cm,height=8cm]{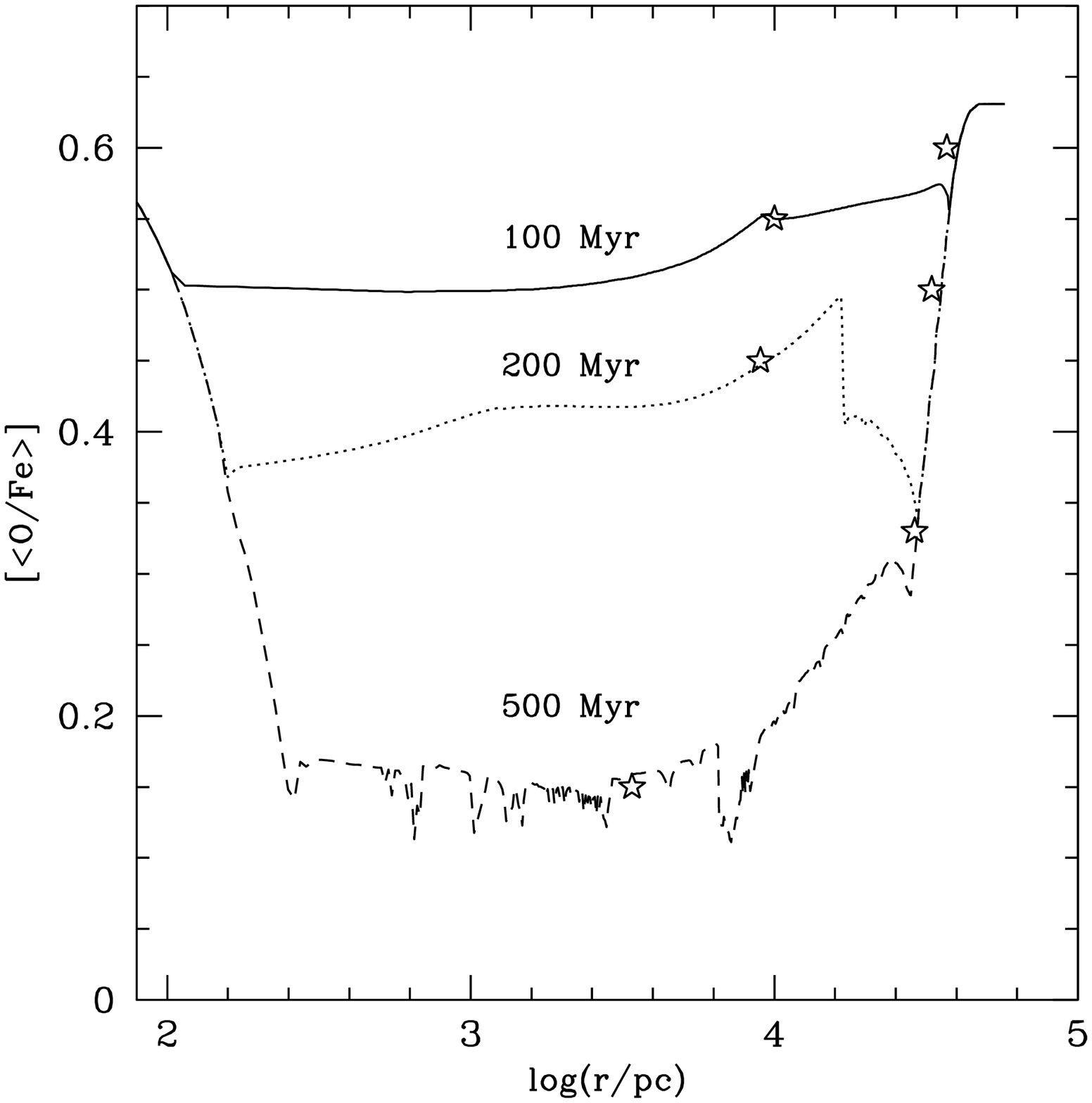}
\caption{Time evolution of radial [$<O/Fe>$] gradient in stars predicted by model La
at different times (only mass-weighted values)
Solid: 100 Myr; dotted: 200 Myr; dashed: 500 Myr.  }
\label{grad_evol2}
\end{figure}

In this section we discuss the issue of radial
gradients in the stellar abundance ratios. We concentrate on the
\emph{actual} gradients, namely on the ones whose properties 
can be measured by an observer.  A snapshot of model La after 100
Myr, reveals gradients already in place with slopes
$\Delta_{O/Fe}=0.08$ (Fig.~\ref{grad_evol2}) and $\Delta_{Fe/H}=-0.35$ 
(luminosity-weighted, upper panel of Fig.
\ref{grad_evol}).
After the SF has been completed, we have $\Delta_{O/Fe}=0.19$ and
$\Delta_{Fe/H}=-0.5$, respectively. Both values are consistent with
the predictions by PM04.  In the same time interval, $R_{core,*}$ and
$R_{eff,*}$ decrease by a factor of 3 and 1.5, respectively.  The
changes in these quantities are more evident if we look at 
other models such as 
Ma1, where the
final $R_{core,*}$ and $R_{eff,*}$ are smaller by a factor of 5 and 2
than the \emph{initial} ones, respectively. In this case, however, the
slope in the [$<O/Fe>$]
changes more smoothly from -0.024 to 0.02,
whereas the steepening in the Fe gradient (from 0.48 to -0.13) is more dramatic.

In conclusion, both models Ma1 and La experience an outside-in
formation process, which creates the abundance gradients, within the observed
range, although with different slopes.  At this stage we can say that
the galactic winds certainly play a role in the gradients build-up.
 The temporal evolution of the gradients for model La can be visualized in
Fig.~\ref{grad_evol}, where the mass-weighted values for the
[$<Fe/H>$] are also displayed in the bottom panel. As expected from the 
analysis of PMC06, mass-weighted values might differ from luminosity-weighted quantities
with increasing galactocentric radii, owing to the well-know strong
metallicity dependence of the light in the optical bands.  
In this particular case, we predict either a  quite flat 
gradient, when the mass-weighted values are taken into account.
This happens because also at large radii there is a significant number
of very metal-rich stars, even though the peak of the stellar metallicity distribution
(see Fig.~\ref{gdwarf_1}) occurs at a lower [Fe/H] with respect to the
core. There are many concurring effects which generate this apparent dichotomy
between peak values and averages. First of all, we remind the reader
that the stellar metallicity distributions are generally asymmetric, thus
the mathematical average does not coincide with the distribution's \emph{mode} (i.e.
the peak value, see PMC06). Secondly, the integral in eq.~\ref{PP75} 
is performed by taking into account a linear sampling of star mass in Fe/H bins (instead of [Fe/H]).
In other terms, $[<Fe/H>]$ is always higher than  $<[Fe/H]>$ (see Gibson 1997).
Therefore we stress that taking the observed (i.e. luminosity-weighted)
gradients at their face values, it might not necessarily reflect the
actual galaxy formation process. Moreover all these subtle differences
in the choice of a SSP-equivalent value (either $[<Fe/H>]$ or $<[Fe/H]>$ or simply
$[Fe/H]_{peak}$) might lead to different final value for our gradients.

In order to guide the eye, in the upper panel of figure 6 the solid 
lines represent a linear
regression fit of the mean 
(luminosity weighted) abundances, at each time, at the core and 
at the effective radius.  With this example we want to give
a warning: if an observer measures
the abundance at both $R_{core,*}$ and $R_{eff,*}$ and then tries to infer a 
metallicity gradient by a linear regression (i.e. a straight line of slope 
$\Delta_{Fe/H}$), the difference between its findings and the actual behaviour 
of [$<Fe/H>$] versus the radius can be large.

%In general we have that, while the chemical properties of the
%outermost regions slowly evolve in time, the innermost
%ones (at $\sim R_{core,*}$) have a strong increase in the Fe locked-up
%in stars and a consequent decrease of the $\alpha$-enhancement.  The
%reasons of such a strong evolution of the central value of the
%[$<O/Fe>$] abundance ratio will be further addressed in the next
%sections.  

By means of these models we have shown that a 10\% SN
efficiency, as adopted in purely chemical models (PM04, PMC06,
Martinelli et al. 1998), is supported also by hydrodynamical models.
In passing, we note that models with 100\% SN efficiency (e.g. MaSN)
undergo the galactic wind too early in their evolution, thus implying
ithat their chemical properties are at variance with
observations.

\subsection{Gradients in Fe/H and total metallicity}

We find a radially decreasing luminosity-weighted Fe abundance in all our models:
$\Delta_{Fe/H}$ spans the range -0.5 -- -0.2 dex per decade in radius,
with a mean value of -0.25, in good agreement with the analysis of
Kobayashi \& Arimoto (1999).  Once transformed into observables by
means of 12 Gyr old TMB03 SSPs, the predicted gradient slopes are $d
Mg_2/log (R_{core,*}/R_{eff,*})\sim -0.06$ mag per decade in radius,
again in agreement with the typical mean values measured for
ellipticals by several authors  and confirming the PM04 best
model predictions.  We notice that for models such as Mb3 and Ma2, we
obtain $d Mg_2/log (R_{core,*}/R_{eff,*})\sim -0.1$ mag per decade in
radius, possibly matching a few objects in the sample of
Ogando et al. (2005, see also Baes et al. 2007).  
%\footnote{The analysis of possible correlation between
%the predicted slopes and the galactic mass will be postponed to a
%forthcoming paper.}  
This conclusion is strenghtened by the fact that
also the \emph{total} metalliticy gradients, are similar among all
the models, their slopes typically being $d([<Z/H>_V])/log
(R_{core,*}/R_{eff,*})\sim -0.2 - -0.3$ dex per decade in radius,
in agreement with the average value of the
Annibali et al. (2006) sample, with the remarkable exception of model
Ma2 (an average elliptical with
the gas initially in equilibrium at $10^4$ K - as well as $\epsilon_{SF}$=10)
whose slope  of $d([<Z/H>_V])/log (R_{core,*}/R_{eff,*})$=-0.42 dex per
decade in radius is close to the largest gradients observed in the
galaxies in the sample of Ogando et al. (2005).

The build-up of such gradients can be explained 
to the non-negligible role of the galactic wind, which occurs later
in the central regions, thus allowing a larger chemical enrichment
with respect to the galactic outskirts.
{ The predicted gradient slopes are independent from the choice of the intial setup given by either
case \emph{a} or \emph{b}.}
We are conscious, however, that we relaxed the PM04 hypothesis of 
not-interacting shells; therefore, in the rest of the paper we will also 
highlight the role of the metal flows toward the center.

\subsection{Gradients in O/Fe}

Recent papers as Mehlert et
al. (2003), Annibali et al. (2006) and Sanchez-Blazquez et al. (2007)
have shown a complex observational situation 
relative to abundance gradients, especially the gradients of the [$\alpha$/Fe] ratio. 
A successful galactic model should be able to
reproduce the [$\alpha$/Fe] radial stellar gradient, either if flat or
negative, while keeping fixed all the other properties (including the
[$<Fe/H>_V$] gradient).  This is nearly impossible with standard
chemical evolution codes, unless by using extreme assumptions which
may worsen the fit of all the other observables.

The hydro-code presented in this paper helps us in tackling this
issue.  From the entries in Table 2, in fact, we notice the all the
objects which present reasonable values for their chemical properties,
including the [$<Fe/H>_V$] gradient, show a variety of gradients in the
[$\alpha$/Fe] ratio, either positive or negative, and one model shows no
gradient at all (Mb2, namely an average elliptical with
the gas initially diffuse and cold - $10^4$ K - 
as well as $\epsilon_{SF}$=10).  

\begin{figure}
\includegraphics[width=8cm,height=8cm]{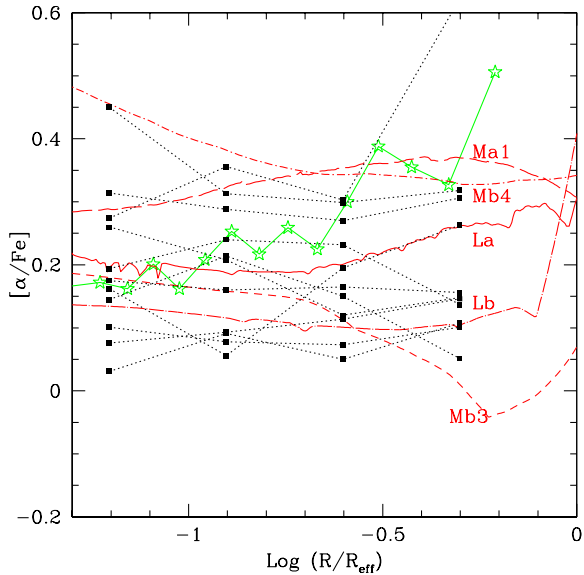}
\caption{[$<O/Fe>$] gradients predicted by several models compared to a subsample of Annibali et al. (2006)'s data
(full squares connected by dotted lines) and the data for NGC4697 (stars).}
\label{confronto}
\end{figure}

{ A comparison between some of our models and data drawn from Annibali et al. (2006) paper (namely a subsample
of only massive ellipticals with homogeneously measured gradients out to $0.5 R_{eff}$) is made in fig.~\ref{confronto}.
Also the data for NGC 4697 are reported.}
The models predict a relationship between 
the abundance ratios and the radius which is not linear.
This further complicates the
comparison with observations and will be the subject of a future paper.  As
an example we only notice that the Annibali et al. (2006) sample is limited
to $\sim R_{eff,*}/2$; therefore, it is not surprising that their
mean slopes are smaller than expected if
one takes into account the whole region $R_{core,*}\le R \le R_{eff,*}$. 
However the agreement with our model is very good, when considering the same
galactic regions, see fig.~\ref{confronto}

As expected from this comparison, the predicted values for $\Delta_{O/Fe}$  span a range from  -0.2 to + 0.3, which is similar 
to the observed one (e.g. Mehlert et al. 2003), whith an average gradient slope of -0.002 dex per decade in radius.

Remarkably, this occurs in spite of the fact
that the galaxy formation process always proceeds outside-in. 
%The only
%exceptions are: model Mb5 (an average elliptical with
%the gas initially diffuse and cold - $10^4$ K - as well 
%as $\epsilon_{SF}$=2) and model MaSN (as Ma1 but with $\epsilon_{SF}$=1), 
%which should be discarded because their
%predicted final radii are too large.

No correlations between $\Delta_{O/Fe}$ and other galactic properties are
found, as expected from observations. We only notice that the
galaxies showing the steepest (both positive and negative) [$<O/Fe>$] gradient
slopes,
have also a quite strong radial decrease in the [$<Fe/H>_V$] ratio,
although a quantitative confirmation needs a sample statistically
richer than ours.  A correlation in this sense seems to emerge by the
Annibali et al. (2006) data (Annibali, private communication).

For instance, in the case of model Ma2, we predict
$\Delta_{O/Fe}=-0.21$, but it has basically the same final stellar
mass of both models La and Lb, being only more compact, and it shows
average abundance ratios in stars matching the typical mean values
observed for massive ellipticals.  On the other hand, model Mb4 (as Ma1, the
only difference is that the gas is diffuse at the beginning) has a
gradient of  $\Delta_{O/Fe}=-0.08$ and model Mb1 predicts
$\Delta_{O/Fe}=0.11$.  We stress again that all these models halt the
SF at $R\ge R_{eff,*}$ earlier than in the core. 

\section{Possible explanations for the observed variety of 
[$<O/Fe>$] radial gradient slopes}

We have analysed the possible causes for the variety of the predicted 
gradients 
in the abundance ratios:
metal-enhanced radial flows and variable timescale of SF with radius.
Here we disentangle their different roles by studying the effects
of the gas flows in determining the central values (i.e.
the [$<O/Fe>_{core,*}$]), whereas the variation
of the SF timescale along the radius will be mainly linked
to the gradient slope.
{ We stress that the results we present here and their interpretation
is valid for the particular initial conditions that we explored.
Therefore, the intial set-up of the simulations is a sufficient
condition for such a variety of gradient to be created}.

\subsection{Radial gas flows}

In order to understand the differences - both observed and predicted - in 
the [$\alpha$/Fe] gradients among ellipticals, we first study the gas 
composition in a sphere of radius $R_{core,*}$ at each time-step.
In this way we can have insights on the role of the gas
flows in the determination of [$<O/Fe>_{core,*}$].

Almost all the models predict that, after the first 100 Myr, 
a substantial fraction (i.e. 80-90\%) 
of the metals present in the gas inside $R_{core,*}$ has an external { (i.e. $R_{core,*} < R < R_{tidal,*}$)} origin.
This means that a non-negligible contribution to the gradients is 
due to the gas flows, as shown by the negative velocity field for
$t<t_{gw}$ in Figs.~\ref{hydro_q} and \ref{hydro_qqtr}, 
and this is also expected in dissipative models such as the Larson (1974) and the Carlberg (1984) 
ones.
This effect cannot be seen in standard chemical evolution models with 
non-interacting shells,
where, at a fixed mass, the predicted $\Delta_{Fe/H}$ is always smaller than
in the models presented in this paper (e.g. see Table 5 of PM04).

%However, to better understand the different gradients in the $\alpha$ to Fe 
%ratios, we need a more meaningful quantity, which tells us the ratio 
%of the speed of star formation rate (i.e. the time derivative of the mass locked into stars) 
%to the gas inward flux (the time derivative of the mass flowing to the galactic centre)
%which supplies with metals produced in the galactic outskirts.
%If we compare the above quantities by means of the ratio between the inflow and
%the star formation timescales, we find that in all the studied cases it is lower than unity for most of
%their lifetime, although with relative differences among the models up to
%a factor of ten.
%This means that, in general, the inflow rate is higher than the star formation rate,
%but in models such as Lb the above mentioned ratio is around 0.8 (and, remarkably  higher than 1 in the first few Myr)
%, whereas in models such as Ma1 or Mb4 can be as low as 0.3.
%In these latter cases, we expect a stronger contribution
%of the external regions in injecting metals in the galactic core,
%with respect to model Lb.

In order to quantify the effects of the convolution of the SF with
the gas flows, we make a step further and use eq.~\ref{PP75}
in order to define, for a given chemical element, the mass-weighted ratio ${\cal R}$ between the mass of this element
produced in the galactic core and locked-up in stars to the amount produced in a more external region
(and subsequently locked-up in stars inhabiting the core). 
In particular, for O we define the quantity ${\cal R}_O$ as: 
\begin{equation}
{\cal R}_O={1\over S_{f,core}} \int_0^{S_{f,core}} (O_{out}/O)(S) \,  dS\, ,
\label{ro}
\end{equation}
{ where, at variance with eq.~\ref{PP75}, we now consider the distribution of
stars as function of the ratio $(O_{out}/O)$ and extract its average.
In this case $O_{out}$ is the mass of O
produced in the external (i.e. outside $\sim 3\times R_{core,*}$) 
part of the galaxies, sunk in the galactic centre because of the 
radial inflows and eventually locked-up in
stars inhabiting the core \footnote{We only
subtract from the metal budget, a posteriori, those elements not
produced in situ and followed in their evolution by means of a suitable
tracer}.
On the other hand, O is the \emph{actual} mass of oxygen out of which
stars form inside a sphere of radius $R= R_{core,*}$.  
A high efficiency of the radial flows in transferring
O from the external regions of the galaxy will correspond to high values of the ratio ${\cal R}_O$.
On the other hand, ${\cal R}_O = 0$ means that all the stars formed in the
core incorporated only the O produced by the previous generations which
populated the core.}
%Here we consider the final
%values of $R= R_{core,*}$, because we want to estimate the relative
%contribution of the metals produced in the outskirts to the final
%abundance in a given grid point, at variance with Fig.~\ref{grad_evol2}
%where we show the \emph{actual} build-up of the gradient as seen by an
%observer.
%In this case the integration of eq.~\ref{ro} is performed
%over the mass of stars formed inside a sphere of radius $R= R_{core,*}$
%and, typically,{ $S_{f,core}\sim 0.05\times S_f$ and not at each grid-point.}
%The ${\cal R}_O$ ratio basically gives a stellar mass-averaged flux of metals produced in the outermost
%regions, which contributes to the central value of [$<O/Fe>$].
We also evaluate the same ratio in the case of the Fe, namely
${\cal R}_{Fe}$. Both the ${\cal R}_O$ and the ${\cal R}_{Fe}$ time evolutions for four
selected models are shown in Table 3.  These quantities give an estimate 
of the contribution of the metal rich radial flows to the
build-up of the gradients.  The last row of Table 3 shows the quantity
[$<O/Fe>_{*,core,noflux}$], namely the expected central value of the
[$<O/Fe>_{*}$] in the hypothetical case in which the metals produced
outside $\sim 3\times R_{core,*}$ do not flow into the core (to be compared
to the entries of Table 2, 6th column). 
%We stress that these results are obtained without changing the hydrodynamical
%evolution: we do not halt the gas flows.  
%We only
%subtract from the metal budget, a posteriori, those elements not
%produced in situ and followed in their evolution by means of a suitable
%tracer. 
%In practical terms it means that the integral in
%eq.~\ref{PP75} is performed only taking 
%into account the O and the Fe
%produced inside $\sim 3\times R_{core,*}$.  The gas velocity profile
%of all models guarantees that these metals do not move out of $\sim
%3\times R_{core,*}$ radius sphere while the SF is active.

\begin{table*}
\centering
\begin{minipage}{120mm}
\scriptsize
\begin{flushleft}
\caption[]{Contribution to the core abundances by metals produced in more external regions}
\begin{tabular}{l|llll|llll}
\hline
\hline
t (Gyr) &   &${\cal R}_O$ & &&& ${\cal R}_{Fe}$&                             \\
        & Ma1 &Mb4    & Mb3 & Lb &       Ma1 &Mb4    & Mb3 & Lb\\
0.05    &0.49 &0.66  &0.82  &0.23&      0.49 &0.66&  0.82& 0.21\\
0.07    &0.70 &0.75  &0.84  &0.45&      0.69 &0.74&  0.84& 0.42\\
0.10    &0.81 &0.82  &0.85  &0.58&      0.80 &0.82&  0.86& 0.56\\
0.15    &0.86 &0.79  &0.78  &0.67&      0.85 &0.82&  0.83& 0.67\\
0.20    &0.88 &0.76  &0.72  &0.64&      0.87 &0.82&  0.80& 0.73\\
0.30    &0.89 &0.76  &0.64  &0.57&      0.89 &0.81&  0.74& 0.72\\
final   &0.89 &0.76  &0.64  &0.57&      0.89 &0.81&  0.74& 0.72\\
\hline
[$<O/Fe>_{*,core,noflux}$]&0.27&0.51&0.33&0.30& \\
\hline
\end{tabular}
\end{flushleft}

\end{minipage}
\end{table*}

Remarkably the 3/4 of the models have
basically [$<O/Fe>_{*,core,noflux}$]=0.3.
%Therefore, we expect that the gas flows can affect the central
%values and, thus, change the gradient slope.

In model Ma1, the mild positive $\Delta_{O/Fe}$ is not enhanced by the
metal rich gas produced in the ouskirts and flowing toward the
center because $ {\cal R}_{Fe}$ and $ {\cal R}_O $ evolve in lockstep
because they are dominated by the external production of metals
in the same way (see Table 3).
%, even though
%they exhibit very high values. 
%This is due to the convolution
%of the SF rate with the fact that at every
%timestep the gas coming from the outskirts has basically the same (high) [O/Fe]
%abundance ratio of the gas already present in the galactic core.
%In the galactic centre, in fact, most of the gas supply necessary
%to feed the star formation process is already in place,
%and thus the majority of the stars can form at a rather constant speed

%\begin{figure}
%\includegraphics[width=8cm,height=8cm]{pipino8.eps}
%\caption{The SF rate in the galactic core for two particular cases:
%model Ma1 (\emph{solid line}); 
%model Mb3 (\emph{dotted line}).}
%\label{sfr}
%\end{figure}

In other cases, such as the model Mb3 (average elliptical with
the gas initially diffuse and hot - $10^6$ K - as well as 
$\epsilon_{SF}$=10), instead, we have $ {\cal R}_{Fe} > {\cal R}_O $.
This model, in fact, starts from a uniform gas distribution,
then most of the gas out of which the stars
form, must have first sunk into the centre.
As a consequence, the star formation rate peaks later
with respect to model Ma1.
%(see Fig. \ref{sfr}, dotted line) with respect to model Ma1 (see Fig. \ref{sfr}, solid line).
On the other hand, the outermost regions halt their
star formation process, and thus the O production, quite soon; therefore,
most of the stars in the central regions preferentially
lock the Fe which is coming from the SNIa exploding in the outskirst,
rather than O.
This explains the slightly low central [$<O/Fe>$], despite the high SF 
parameter ($\epsilon_{SF} = 10$);  
%On the other hand, the efficient SF process
%keeps the final values of $ {\cal R}_{Fe}$ and $ {\cal R}_O $ among the lowest in
%our galaxy sample (AGAIN NOT CLEAR!!).  
in fact, $\Delta_{O/Fe}$ is $\sim$ 0.18 at 100 Myr,
$\sim$ 0 at 240 Myr and becomes negative soon after.  

As anticipated above, due to the very fast star formation rate (with
respect to the gas inflow rate) in model Lb we have the lowest ${\cal R}_{Fe}$ and ${\cal R}_O$,
therefore the gradient could reflect the real outside-in formation in
a manner which resembles the multi-zone chemical evolution models with
non-interacting shells.  
Nevertheless, also in this case, we have ${\cal R}_{Fe} > {\cal R}_O$,
which is the outcome of a differential inflow, as explained
above for model Mb3.

%Concerning model Mb4 (as Mb3, but the gas is denser, whereas
%the SF parameter is lower than the latter case), it evolves in a slightly 
%different manner, because the core is already in place at 100 Myr, 
%while the rest of the
%galaxy undergoes a very fast outside-in formation. This hampers a
%substantial contribution in metals from the outskirts and explains why
%the central [$<O/Fe>$] is 0.42 dex and $\sim$0.1 dex higher than the
%outskirts.  We also note that this kind of radial flows helps in
%keeping the predicted central [$<O/Fe>$] within the limits of the
%observed range in models with steep $\Delta_{Fe/H}$. 

In summary, radial flows may lower { the core value of [$<O/Fe>$], (that we consider as} 
the zero point of the gradient in [$\alpha$/Fe])
relative to the case with no radial flows. The reason for that is that $\alpha$-depleted material 
flows from the outermost into the innermost regions. Therefore, the variety of 
gradients 
(in particular positive, null or negative) depends on the efficiency of 
the $\alpha$-depleted gas to flow 
from the outside to the inside during the time of active star formation. 
In other words, it depends on the velocity of the inflowing gas.
Clearly a larger or smaller parameter of SF can have a strong influence 
on this process.
In order to help the visualization of such a complex process,
we show Fig.\ref{figura_esplicativa} where the solid line at the top
represents a hypothetical \emph{pure outside-in model} with non-interacting
shells  and [$<O/Fe>_{*,core,noflux}$]=0.3. { The gradient slope is chosen
to be 0.15 dex per decade in radius. In this case the [$<O/Fe>$] gradient
is set by the occurrence of the galactic wind, which happens earlier
in the outermost regions.
By no means such a model is real. It just helps in visualising the simplest
scenario - which is quite a common assumption in the literature involving
multi-zone chemical evolution modelling - and the differences introduced
by taking into account radial flows and the local variation in the
\emph{input} star formation timescale. None of the models run 
correspond to this ideal case, therefore we cannot compare it
with any of our predicted curves.}

In order to take into account 
the role of the gas flows we then correct the predicted gradient (solid line in the middle), thus
obtaining something similar to the predictions by models La (dashed line) 
and Lb.
This mechanism helps also in explaining the value for [$<O/Fe>_{*,core}$] 
predicted by other models, such as Mb3 (dotted line).

\subsection{The role of the star formation timescale at different radii}

Let us examine now the effect of varying the SF timescale.
The analysis of the mass-weighted abundance ratio in the
inner zone is not enough to explain the gradients. In fact we have
studied only the build-up of the zero-point value, taken as the
quantity [$<O/Fe>_{*,core}$].  Even in the simplistic assumption in
which the gradient can be well represented by a straight line we need
another quantity in order to fix the slope steepness.  We chose to
study the radial variation of $t_{cool}$ and $t_{ff}$, because
another important difference with respect to PM04 and PMC06 is that
here $\nu  = {\epsilon_{SF} \over max(t_{cool},t_{ff})} \ne const.$

We find that $\nu (core) \sim 2-3 \times \nu (R_{eff})$ in a model such
as Lb, which closely follows the PM04 best model.  On the other hand,
models with either a zero or a negative $\Delta_{O/Fe}$ have
$\nu (core) \sim 5-10 \times \nu (R_{eff})$ for most of their evolution,
thus favouring a higher [O/Fe] ratio in the stars belonging to the
inner regions.  
Other test cases, not presented here, show us that if we 
run models with even higher values for the star formation parameter 
(i.e. $\epsilon_{SF}\ge 20$,) the strong feedback by SNe halts the
gas flows; therefore the supply of baryons for the SF in the galactic 
center is strongly reduced and the outcome
is a too diffuse galaxy. A similar result can be obtained by increasing
$\epsilon_{SN}$, as shown by model MaSN.

\begin{figure}
\includegraphics[width=8cm,height=8cm]{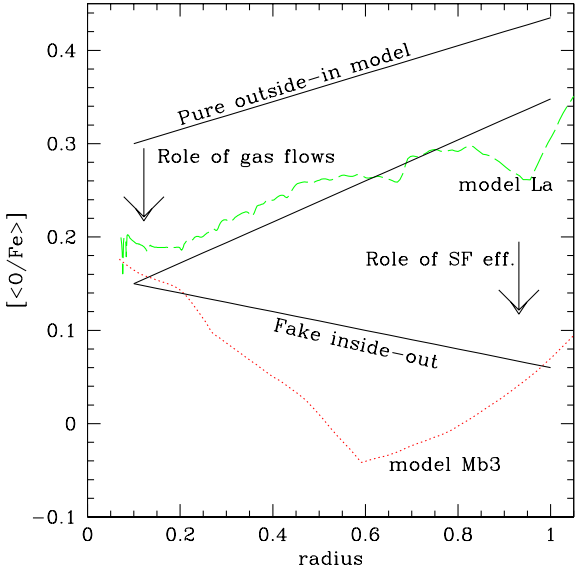}
\caption{A sketch of the relative contribution of the gas flows strenght and 
the star formation parameter $\nu $
to the creation of the final gradient for two particular cases:
model La's predicted gradient (positive slope,\emph{dashed line}); 
model Mb3's predicted gradient (negative slope, \emph{dotted line}).
\emph{Pure outside-in model:} hypothetical model with an outside-in formation, 
[$<O/Fe>_{*,core,noflux}$]=0.3 (no gas flows) and $\nu $ constant with radius;
\emph{fake inside-out model:} hypothetical model with a strong variation of the star formation timescale
with radius.
The abscissa is expressed in units of the effective radius.}
%\emph{fake inside-out model:} hypothetical model with a strong variation of the star formation efficiency.}
\label{figura_esplicativa}
\end{figure}

The radial variation of $\nu  $ means that the effect of the outside-in formation
could be balanced by the interplay between local differences in the SF
timescale and differential gas flows.
Therefore the combined effect of gas flows plus   
a strong variation in the star formation timescale
along the radius, make the \emph{hypothetical outside-in model} gradient change slope 
(line labelled as
\emph{fake inside-out model} in Fig.\ref{figura_esplicativa}), thus matching the
average trend predicted by model Mb3 (dotted).

In general, $\Delta_{O/Fe}$ seems to fluctuate around a null value
and to be a result of the interplay of many hydrodynamical factors,
which render it more sensitive to the initial conditions of the gas
rather than an indicator of the chemical enrichment process. 
Possible connections between the above mentioned
trends and the other galactic properties will be investigated
in a future paper and will help in sheding more
light on this subject.

In the end we notice that the gradients in $[<Fe/H>_V]$
and $[<Z/H>_V]$ may be affected by the particular formation history of
one model only in their zero point, whereas their slopes are shaped by
the strong role of the gas flows (both the final values of $ {\cal R}_{Fe}$
\emph{and} $ {\cal R}_O $ are always larger than the 50\%) and by the fact
that the SF proceeds always outside-in.

%On top of this, we add the warning that recent (small) episodes of SF in the center
%can futher perturb the gradients, by yielding both smaller ages and average abundance
%ratios in the stars biased by the youngest SSPs. Therefore, if they are real they
%can contribute to some degree to the scatter in the observed slopes.

%Finally, we also predict a flattening of the gradients at $R \le
%R_{core,*}$ (e.g. Fig. \ref{grad_evol}), but for the very first gridpoints , 
%which are irrelevant,
%given the limited number of stars formed in such small regions.  
%More interestingly, in the region $R_{core,*}\le R \le R_{eff,*}$ the
%gradient slope is a function of the radius (as already showed by PM04,
%compare the entries of their Table 5).
%In fact, as it can be clearly seen in Figs. \ref{grad_evol}, \ref{grad_evol2}  
%and \ref{figura_esplicativa},
%in many cases the models predict a relationship between 
%the abundance ratios and the radius which is not linear.

Another important point is that the differences among the values of 
$\Delta_{O/Fe}$ in the models presented in this paper are typically around 
a factor of 2 (if they
are not presented in logarythmic units), values which are probably
comparable to all the uncertainties involved in the measurements of
the gradients as well as uncertainties related to the transformation 
from indices to abundances of such (see PMC06). 
This fact calls for newer, larger as well as
homogeneous samples of gradients observed in ellipticals and extended to
one effective radius. Only then, in fact, it will be possible to 
discriminate among the particular models presented in this work.

\section{Conclusions}

In this paper we have studied the formation and evolution of
ellipticals by means of hydrodynamical models in which we implement
detailed prescriptions for the chemical evolution of H, He, O and Fe,
thus presenting a
quite detailed treatment of both the chemical and the gas-dynamical
evolution of elliptical galaxies.  Within this framework we are able
to relax the assumption of non-interacting shells which hampers many
chemical evolution codes in the modelling of the gas flows, thus allowing 
us to perform a detailed  study of the build-up of
the metallicity gradients in stars.  We suggest
an outside-in formation for the majority of
ellipticals in the context of the SN-driven wind scenario, thus confirming previous results of chemical evolution models, but we also show the necessity of taking into account in detail gas inflows/outflows. 
%The outside-in formation is due to the fact that ellipticals stop
%forming stars in the outermost regions before than in the innermost
%ones.

Here we summarise our main results.

\begin{itemize}
\item We find $\Delta_{Fe/H}$ in the range -0.5 -- -0.2 dex per decade
in radius and $\Delta_{Z/H}\sim$ -0.3 dex per decade in radius, in
agreement with the observations (e.g. Kobayashi \& Arimoto, 1999).

\item These gradients in the
abundances, once transformed into
predictions of the line-strenght indices, lead to 
$d Mg_2/log (R_{core,*}/R_{eff,*})\sim -0.06$ mag
per decade in radius, again in agreement with the typical mean values
measured for ellipticals  and confirming the
PM04 best model predictions.  We also find that some models predict 
a steeper gradient, and this seems to be in agreement with the observed 
gradients of a few massive
objects in the Ogando et al. (2005) sample.

\item The build-up of the gradients is very fast and we predict
negligible evolution after the first 0.5 - 1 Gyr.

%\item Most of the galactic models predict that stars form outside-in, as 
suggested by PMC06.

\item We also find the \emph{actual} (i.e.
mass-averaged) metallicity gradients can be flatter than the
luminosity-weighted ones (i.e. the observed ones).

\item The main novelty of this work is that we address the issue of the 
observed scatter in the radial
gradient of the mean stellar [$\alpha$/Fe] ratio and its apparent lack of
any correlation with all the other observables. 
%This problem cannot be
%tackled with the state of the art chemical evolution models.  
By analysing typical massive ellipticals, we find that all the models
which  predict chemical properties, including the
[$<Fe/H>_V$] and the global metallicity gradients, within the observed
ranges, show a variety of gradients in the [$\alpha$/Fe] ratio, either
positive, negative  or null.

\item We explain this finding with the fact that the suggested outside-in
mechanism for the formation of the ellipticals is not the only process
responsible for the formation of abundance gradients.  In particular,
other processes should be considered such as the interplay between
local differences in the SF timescale and gas
flows. In particular, our models suggest the gradient in the [$\alpha$/Fe]
ratio to be be related to the interplay between the velocity of the 
$\alpha$-enhanced radial flows, moving from the outer to the inner galactic 
regions, and the intensity and therefore duration of the SF formation process 
at any radius.
In other words, if the flow velocity is fast relative to the star formation, 
the stars still forming at inner radii have time to form out of 
$\alpha$-enhanced gas coming from the outermost regions, thus flattening and 
even reversing the sign of the [$\alpha$/Fe] gradient.

\item { In particular, we have shown that we do not need the merger events 
invoked in order to have a shallow [$<\alpha/Fe>$]
gradient.}

\item Moreover, the predicted age gradients
are very small, being typically a few Myrs per decade in radius, in agreement
with Sanchez-Blazquez et al. (2007). This
means that the estimate of the \emph{relative} duration of the SF
process between two different galactic regions by measuring the
[$<O/Fe>$] is not a robust method.  In other cases in which, instead,
the age gradients are stronger we expect a much more evident radial
variations in the [$<O/Fe>$], as those outlined in PMC06.

\item According to our fiducial cases, up to the 90\% of the metals locked 
in stars 
in the galactic center could have been synthesized at larger radii.

\end{itemize}

We stress that the new class of models presented here make several
and new predictions on both the shape and the \emph{fast} evolution of the metallicity gradients
which are left unconstrained by the lack of observations.
What makes galaxies start from quasi-monolithic conditions is still
to be understood. The quest for an explanation of such behaviours
will be a challenging field
of research if future observational campaigns will confirm
the steep positive $\Delta_{O/Fe}$ found in, e.g., NGC 4697
(see fig.~\ref{confronto}), thus, further validate a particular type
of models.
These observables will be the testbench
for our suggested galaxy formation scenario to be tested by future observations.

%\acknowledgments
\section*{Acknowledgments} 
We acknowledge useful discussions with F.Annibali, L.Ciotti.
 
Then we warmly thank F. Calura, C. Chiappini, S.Recchi and P. Sanchez-Blazquez for a careful 
reading of the paper and many enlightening comments.
The work was supported by the Italian Ministry for
the University and the Research (MIUR) under COFIN03 prot. 2003028039.

\clearpage


\begin{thebibliography}{}
\bibitem[]{}Annibali, F.; Bressan, A.; Rampazzo, R.; Zeilinger, W. W.; Danese, L., 2007, A\&A, 463, 455
\bibitem []{}Arimoto, N., $\&$ Yoshii, Y. 1987, A$\&$A, 173, 23 
\bibitem[]{}Asplund M., Grevesse N., Sauval A. J., 2005, ASPC, 336, 25
\bibitem []{}Baes, M., Silchenko, O.K., Moiseev, A.V., Manakova, E.A., 2007, A\&A, 497, 991	
\bibitem []{}Bedogni, R. \& D'Ercole, A., 1986, A\&A, 157, 101
%\bibitem []{}Bekki, K., $\&$ Shioya, Y. 1999, ApJ, 513, 108
\bibitem []{}Bower,  R.G., Lucey,  J.R., Ellis,  R.S. 1992, MNRAS, 254, 589
\bibitem[]{} Carlberg, R.G., ApJ, 286, 403
\bibitem []{}Carollo, C.M., Danziger, I.J., $\&$ Buson, L. 1993, MNRAS, 265, 553
\bibitem []{}Ciotti,  L., D'Ercole,  A., Pellegrini,  S., Renzini,  A. 1991, ApJ, 376, 380
\bibitem []{}Davies, R.L., Sadler, E.M., $\&$ Peletier, R.F., 1993, MNRAS, 262, 650
\bibitem []{}Forbes, D.A., Sanchez-Blazquez, P., Proctor, R., 2005, MNRAS, 361, 6
%\bibitem []{}Faber, S.M., Worthey, G., $\&$ Gonzalez, J.J. 1992, in IAU Symp. n.149,
%eds. B. Barbuy $\&$ A. Renzini, p. 255
\bibitem []{}Friaca, A.C.S.; Terlevich, R.J. 1998, MNRAS, 298, 399
\bibitem []{}Gibson, B.K., 1996, MNRAS, 278, 829
\bibitem []{}Gonzalez, J.J., $\&$ Gorgas, J. 1996, in ASP Conference Series, 86, Fresh Views of Elliptical Galaxies,
eds. A. Buzzoni, A. Renzini, $\&$ A. Serrano, 225
\bibitem []{}Graham, A., Lauer, T.R., Colless, M. \& Postman M., 1996, ApJ, 465, 534
\bibitem[]{} Greggio, L. 2005, A\&A, 441, 1055
%\bibitem []{}Greggio, L., Renzini, A., 1983 in 
%``Frascati Workshop on First Stellar Generations'',  Memorie Societa Astronomica Italiana, vol. 54, p. 311
\bibitem[]{} Harris, W.~E., \& Harris, G.~L.~H.\ 2002, AJ, 123, 3108
\bibitem[]{}Iwamoto, K.; Brachwitz, F.; Nomoto, K.; Kishimoto, N.; Umeda, H.; 
Hix, W. R.; Thielemann, F.K. 1999, ApJS, 125, 439
\bibitem []{}Jimenez, R., Padoan, P., Matteucci, F., Heavens, A.F., 1998, MNRAS, 299, 123
\bibitem []{}Kobayashi, C., 2004, MNRAS, 347, 740
\bibitem []{}Kobayashi, C., Arimoto, N., 1999, ApJ, 527, 573
\bibitem []{}Larson,  R.B. 1974, MNRAS, 166, 585
\bibitem []{}Martinelli, A., Matteucci, F., Colafrancesco, S., 1998, MNRAS, 298, 42
\bibitem []{}Matteucci, F., 1992, ApJ, 397, 32
\bibitem []{}Matteucci, F. 1994, A$\&$A, 288, 57
%\bibitem []{}Matteucci, F. 2001, The chemical evolution of the Galaxy, Kluwer Academic Publishers, Dordrecht
\bibitem []{}Matteucci,  F.,  Greggio,  L. 1986, A$\&$A, 154, 279

\bibitem[]{} Matteucci, F.; Panagia, N.; Pipino, A.; Mannucci, F.;
                    Recchi, S.; Della Valle, M. 2006MNRAS, 372, 265
\bibitem[]{} Matteucci, F.,Recchi, S., 2001, ApJ, 558, 351 
\bibitem[]{} McCarthy, I.G.; Bower, R.G.; Balogh, M.L. 2006 astro-ph/0609314
\bibitem[]{}Mehlert, D.; Thomas, D.; Saglia, R. P.; Bender, R.; Wegner, G.  2003, A\&A, 407, 423
\bibitem []{}Mendez, R.H., Thomas, D., Saglia, R.P., Maraston, C., Kudritzki, R.P., \& Bender, R., 2005,
ApJ, 627, 767
%\bibitem []{}Nomoto, K., Hashimoto, M., Tsujimoto, T., Thielemann, F.K., Kishimoto, 
%N., Kubo, Y., Nakasato, N., 1997, Nuclear Physics A, A621, 467
\bibitem []{}Nelan, J.E.; Smith, R.J.; Hudson, M.J.; Wegner, G.A.; Lucey, J.R.; 
Moore, S.A.W.; Quinney, S.J.; Suntzeff, N.B. 2005, ApJ, 632, 137
\bibitem []{}Ogando, R.L.C., Maia, M.A.G., Chiappini, C., Pellegrini, P.S.,
Schiavon, R.P., da Costa, L.N., 2005 ApJ, 632, 61
\bibitem []{}Padovani, P., \& Matteucci, F. 1993, ApJ, 416, 26
\bibitem []{}Pagel, B.E.J., $\&$ Patchett, B.E. 1975, MNRAS, 172, 13 
\bibitem []{}Peletier, R.F., Davies, R.L., Illingworth, G.D., Davis, L.E., Cawson, M. 1990, AJ, 100, 1091
\bibitem []{}Pipino,  A.,  Matteucci,  F. 2004, MNRAS, 347, 968 (PM04)
\bibitem []{}Pipino,  A., Matteucci,  F., Borgani,  S., Biviano,  A. 2002, NewA, 7, 227
\bibitem []{}Pipino, A.; Matteucci, F.;
                    Chiappini, C. 2006 ApJ, 638, 739 (PMC06)
\bibitem []{}Proctor, R.N., Forbes, D.A., Forestell, A, \& Gebhardt, K., 2005, MNRAS, 362, 857
\bibitem []{}Salpeter, E.E., 1955, ApJ, 121, 161
\bibitem []{}Sanchez-Blazquez, P.; Forbes, D. A.; Strader, J.; Brodie, J.; Proctor, R. 2007, astro-ph/0702572
\bibitem []{}Serra, P. \& Trager, S., 2007, MNRAS, 374, 769
\bibitem []{}Silich, S.A., \& Tenorio-Tagle, G., 1998, MNRAS, 299, 249.
\bibitem []{}Sutherland,  R.S.,  Dopita,  M.A. 1993,  ApJS, 88, 253
%\bibitem []{}Tantalo, R., \& Chiosi, C., 2004, MNRAS, 353, 917

\bibitem []{}Thomas, D., Maraston, C., $\&$ Bender, R., 2003, MNRAS, 339, 897 
\bibitem []{}Thornton,  K., Gaudlitz,  M., Janka,  H.-T.,  Steinmetz,  M. 1998, ApJ, 500, 95
\bibitem []{}Trager, S.C., Faber, S.M., Worthey, G., Gonzalez, J.J., 2000a, AJ, 119, 1654

\bibitem []{}Yoshii, Y., \& Arimoto, N., 1987, A$\&$A, 188, 13 
\bibitem []{}Worthey, G., Faber, S.M., $\&$ Gonzalez, J.J. 1992, ApJ, 398, 69

\end{thebibliography}
\end{document}